\documentclass{article}
\usepackage{graphicx} 
\usepackage{color}
\usepackage{url}
\usepackage{hyperref}
\newcommand{\thepapers}[1]{{\color{blue}#1}}
\newcommand{\available}[1]{{\color{red}#1}}

\newcommand{\hide}[1]{}

\title{Building a Theory of Distributed Systems:  Work by Nancy Lynch and Collaborators}
\author{Nancy Lynch}
\date{March 3, 2025}

\begin{document}

\maketitle

\begin{figure}[h]
\includegraphics[width=\textwidth]{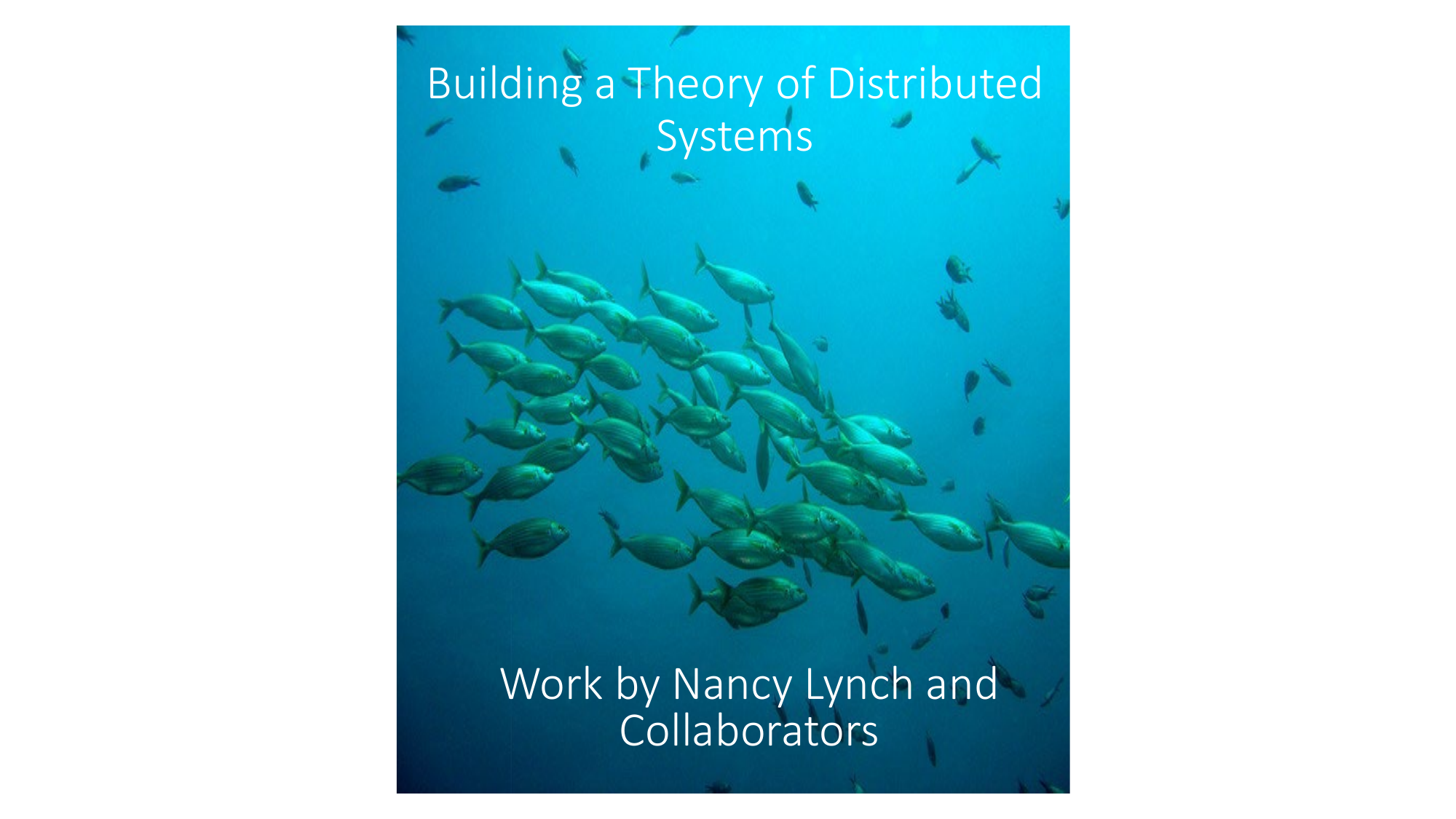}
\caption{If this were a book, this could be its cover design.  The school of fish is a kind of natural distributed system.  The school could also represent all the collaborators.}
\end{figure}

\section{Introduction}

In this manuscript, I summarize research by myself and my very many students, postdocs, and other collaborators, on developing a theory for the field of distributed computing.  
I hope that it provides an interesting look at some of the early work that helped to build this field.

The main theme of our research in distributed computing theory is simply this:
\emph{Distributed systems are everywhere, and we need a usable formal theory to support their design and development.}  
Moreover, a formal theory for distributed computing requires new kinds of models and theoretical techniques, different from those used for sequential computing.
The new theory should support formal description of distributed algorithms and systems, proofs of their correctness, and analysis of their performance.
Also, it should support proofs of impossibility results.
Impossibility results are very hard to prove for sequential algorithms, but they are feasible for distributed algorithms, because the platforms on which distributed algorithms run are much less well-behaved.

\paragraph{Contents of this manuscript:}
Our research in distributed computing theory spans from 1976 until the present (almost 50 years!), and is contained in many hundreds of papers.
To make this manageable, I have selected a small number of "key publications" on which to focus:  25 papers and 3 books.  
In choosing these, I have favored papers that were influential in the field, papers that were influential in our own later work, papers that I was especially invested in, and some other favorites.  I apologize ahead of time for the many other wonderful papers that I have omitted, particularly some excellent papers by my students.
Although I have focused the discussion on these 28 publications, along the way I mention many more.

The manuscript begins with a brief Section~\ref{sec: complexity} describing my background in complexity theory, which is what I worked on before I got interested in distributed computing.
Sections~\ref{sec: gatech} and~\ref{sec: earlyMIT} describe our early work on the beginnings of the field, through 1990, starting at Georgia Tech and continuing at MIT.  My work at Georgia Tech included basic algorithms and lower bounds for shared-memory distributed systems solving basic problems of resource allocation, such as mutual exclusion.  It also included preliminary work on the problem of distributed consensus, which became a very popular research direction for the field, and on general formal models for distributed computing.  My early MIT work involved more research on distributed consensus, including exact and approximate versions of the problem and work based on different timing models.
Formal models were another focus, as well as applications to distributed database transaction processing.

Sections~\ref{sec: laterMITalgorithms} and~\ref{sec: laterMITmodels} describe work at MIT during the later period 1990-2005.  This included two distinct threads of research.  Section~\ref{sec: laterMITalgorithms} describes work on algorithms and lower bounds, including work on communication protocols, on timing aspects of algorithms, on variations of the consensus problem, and on data consistency.  I also wrote my Distributed Algorithms textbook during this time.
Section~\ref{sec: laterMITmodels} describes work on formal models and methods for distributed systems, including work on timed system models, hybrid (continuous/discrete) system models, and probabilistic system models.  Section~\ref{sec: laterMITmodels} also includes some applications of these formal models and methods to some data management algorithms, timed and hybrid systems, and security protocols.

Section~\ref{sec: wireless} summarizes our recent work (after 2005) on algorithms for wireless networks, including algorithms for distributed data management and algorithms built over a Virtual Node abstraction layer.
Finally, Section~\ref{sec: bio} touches briefly on our recent work on biological distributed algorithms, specifically, for insect colonies and brain networks.

\paragraph{Accessing the papers:}
For some help in accessing the key papers, here is a link to a web page that contains a list of the key publications highlighted in this manuscript, with information about where they can be found: \\
\available{https://groups.csail.mit.edu/tds/lynch-papers-highlights.html}. \\
In most cases, I have pointed to the publishers' pages for the final versions of the papers.
For a few papers that I cannot locate easily, or for which we have made significant corrections post-publication, I have included a .pdf.

Throughout, I have referenced many other papers.  Among those, I have highlighted our own papers in red; all of these should be available on-line.

\section{Complexity Theory, 1970-1975}
\label{sec: complexity}

Complexity theory studies bounds on the inherent costs of algorithms for solving precisely-defined problems. In the years when I worked in this field, the studies were almost entirely about sequential algorithms.

I wrote my PhD thesis in the subfield of \emph{abstract complexity theory} in the MIT Mathematics Department, working with Profs. Albert Meyer and Michael Fischer.  The paper~\available{\cite{LMF76}} summarizes my PhD work.  
Briefly, this paper and thesis extended Manuel Blum's axiomatic treatment of the computational complexity of partial recursive functions to relatively computable functions, as computed, for example, by Turing machines with oracles.
My results were quite theoretical, for example, showing that there exist problems with provably high complexity, even when solutions to other problems are given "for free".
The paper went on to study formal reducibilities between problems, defined by complexity bounds.
I did all this in terms of general, abstract complexity measures.

The paper~\available{\cite{LadnerLS74}}, joint with Richard Ladner and Alan Selman, moved away from abstract complexity theory to study \emph{polynomial-time reducibilities}, such as the two defined by Cook~\cite{Cook71} and Karp~\cite{Karp72} respectively.  The point of the paper~\available{\cite{LadnerLS74}} was that many different forms of polynomial-time reducibility had been defined, or could be defined, varying according to the structure of the reducibility.  
The paper~\available{\cite{LadnerLS74}} established the relative strengths of the different reducibilities, with inclusion and separation results. 

So does this work have anything to do with the topic of this manuscript, which is the foundations of distributed computing?
Well, a loose connection is that reducibilities are about solving computational problems in terms of other computational problems.  This is a theme that is important in distributed computing theory, in the form of decomposing complex distributed algorithms into smaller pieces.
In general, working in complexity theory taught me basic techniques of theoretical computer science research.  
It also showed me that I wanted to work on theoretical topics that were more connected with practical problems.
So I soon changed research areas...

\section{Early Explorations of Theory for Distributed Computing, at Georgia Tech, 1976-1981}
\label{sec: gatech}

I started a faculty job at Georgia Tech's School of Information and Computer Science in 1976, where I spent my first year or two exploring new areas of research.  I was looking for places where theoretical computer science methods might apply to practical computing.
Pretty quickly, I gravitated to the new area of distributed computing.

At that time, the Arpanet (the precursor of the Internet) was fairly new, and researchers and developers had begun talking about a new type of computing that they were calling "distributed computing".  At Georgia Tech, I was influenced by Prof. Phil Enslow's work and advocacy for the new area.  
Phil was a retired Army Lieutenant Colonel who had embarked on a second career as a computer science professor.  Phil wrote an interesting position paper~\cite{DBLP:journals/computer/Enslow78} that delineated many features that a distributed system should have.  Basically, he envisioned a high-level, general-purpose programming platform running over a communication network like the Arpanet.
It was supposed to manage distributed data processing, and also to coordinate tasks to be performed on many computers.

It was clear that managing such distributed processing required new algorithms.  The algorithms needed were not traditional single-processor algorithms, but rather, new kinds of "distributed algorithms" that would run on many processors, working in parallel and coordinating with each other.  There was no theory at the time to support the development of such algorithms.

Phil suggested many papers for me to read, notably, papers on distributed database concurrency control.  He understood that I was interested in theoretical aspects of distributed computing, and pointed me to relevant papers by Leslie Lamport and Edsger Dijkstra.
I became very interested in working in the area.

As I began thinking about distributed systems issues, I talked with Mike Fischer at one of the theoretical computer science conferences.  It turned out that Mike was already working on problems in this area, with his PhD student Gary Peterson. Their emphasis was on shared-memory algorithms for mutual exclusion; see~\cite{DBLP:conf/stoc/PetersonF77}.
Mike and I began discussing the area, along with Gary and my (first) PhD student Jim Burns.

Before too long, in around 1978, Mike arranged to take a sabbatical at Georgia Tech in order to work with me and Jim Burns on this topic.
Together we hosted a series of short visits, by Leslie Lamport, Eshrat Arjomandi, Alan Borodin, and others.
At this time, Lamport was already well known for his early work on distributed computing theory, in particular, for the Bakery Algorithm and a collection of algorithms for implementing read/write registers; he had also begun working on the problem of distributed consensus.
He later won the Turing award for this early work.
Also during this time, Prof. Nancy Griffeth joined the School at Georgia Tech; she worked on database algorithms, and also began working with us on distributed resource allocation algorithms.

Overall, I think my years at Georgia Tech were quite productive, in terms of producing concrete theoretical results in the new research area of distributed computing theory.  I discuss some of the highlights of this work in the subsections below.

\subsection{Key publications}

\thepapers{
\noindent
3.1. James E. Burns, Paul Jackson, Nancy A. Lynch, Michael J. Fischer, and Gary L. Peterson. Data requirements for implementation of N-process mutual exclusion using a single shared variable. Journal of the ACM, 29(1):183--205, January 1982. \\

\noindent
3.2.  James E. Burns and Nancy A. Lynch. Bounds on shared memory for mutual exclusion. Information and Computation, 107(2):171--184, December 1993.
Originally appeared in Jim Burns's thesis, around 1981. \\

\noindent
3.3.  Nancy A. Lynch and Michael J. Fischer. On describing the behavior and implementation of distributed systems. Theoretical Computer Science, 13(1):17--43, 1981. Special issue on Semantics of Concurrent Computation. \\

\noindent
3.4.  Eshrat Arjomandi, Michael J. Fischer, and Nancy A. Lynch. Efficiency of synchronous versus asynchronous distributed systems. Journal of the ACM, 30(3):449--456, July 1983. \\

\noindent
3.5.  Michael J. Fischer and Nancy A. Lynch. A lower bound for the time to assure interactive consistency. Information Processing Letters, 14(4):183--186, June 1982.}  

\subsection{Mutual exclusion in shared-memory models}

Our first papers were based on a shared-memory model, in which a number $n$ of processes, operating concurrently and asynchronously with respect to each other, communicate using shared memory to arbitrate exclusive access to "critical regions" of their programs, where they could access a non-shareable resource.
This problem was called the \emph{mutual exclusion} problem.

The setting was very different from standard sequential algorithms, because it involved many active agents acting independently.  
We defined new kinds of models.
Following the general paradigms of sequential algorithms, we gave formal, mathematical definitions of the mutual exclusion problem, gave formal descriptions of efficient algorithms to solve the problem, and analyzed the algorithms.  We also proved corresponding lower bound results.

We studied different fairness guarantees, mainly, a basic \emph{no lockout} property specifying that no process could be prevented indefinitely from reaching its critical regions, and a more stringent \emph{bounded bypass} property that limited the number of times one process could bypass another.

We considered the problem both with powerful \emph{test-and-set} shared memory and with weaker \emph{read/write} shared memory.

\paragraph{Bounds on test-and-set shared memory for mutual exclusion and resource allocation:}
The first paper I was involved in, in the area of distributed computing theory, was Paper 3.1~\available{\cite{BurnsJLFP82}}.
This was based on a shared memory model, in which $n$ processes operated concurrently and asynchronously with respect to each other, communicating using a small shared memory, in order to arbitrate access to critical regions.
To do this, the processes executed two protocols, a \emph{trying protocol} to try to gain access to the critical region, and an \emph{exit protocol} to leave the region gracefully.
This paper defined the mutual exclusion problem, and presented algorithms and nearly-matching lower bounds for variations of this problem.
We considered mutual exclusion with fairness guarantees ranging from no-lockout, to bounded bypass, to simple absence of deadlock (no fairness).

The paper was somewhat influenced by a series of papers by Dijkstra, Knuth, and others, on clever algorithms for mutual exclusion.
However, a difference was that those papers assumed read/write shared memory, whereas
our paper used a stronger type of shared memory, which we called \emph{test-and-set} memory.
In our version of test-and-set, a process could, in one step, read the value of a single shared variable and make some arbitrary changes to both its local state and the shared variable, all atomically.

For me, a stronger influence was a previous unpublished University of Southern California technical report by Armin Cremers and Tom Hibbard.
I learned about this work when I was on the USC faculty together with Cremers and Hibbard, sometime during the years 1973-1976.
Their paper used the test-and-set model.  It showed that two processes cannot achieve mutual exclusion with fairness using a shared variable that can take on only two values.

I found this exciting, because it demonstrated that in distributed computing theory, one can hope to prove nontrivial impossibility results.
This was quite different from the situation in the theory of sequential algorithms, in which lower bound results were (and still are) very hard to prove.
The reason is that distributed algorithms run on very difficult platforms, with many agents, each of which uses only local information.
These difficulties mean that problems are much harder to solve.
The work of Cremers and Hibbard demonstrated how distributed systems might support interesting lower bounds, which later proved to be a key component of work in this area. 

Our Paper 3.1~\available{\cite{BurnsJLFP82}} generalized the earlier work of Cremers and Hibbard to more processes and more types of fairness conditions. \\

\noindent
\thepapers{
3.1. James E. Burns, Paul Jackson, Nancy A. Lynch, Michael J. Fischer, and Gary L. Peterson. Data requirements for implementation of N-process mutual exclusion using a single shared variable. Journal of the ACM, 29(1):183--205, January 1982.} \\

Briefly, we assumed a single shared variable (this restriction doesn't matter for the powerful test-and-set model), and tried to determine the size of a variable that is sufficient to solve the mutual exclusion problem.
It turned out that the answer depends strongly on the type of fairness conditions required.

For instance, a simple $2$-valued semaphore is adequate if we do not require any fairness but just absence of deadlock.
But if we require fairness to all requesting processes, we need some coordination. 
Here is where we encountered the characteristic difficulties of distributed computing with limited communication---how could separate processes, operating independently and asynchronously, manage to coordinate for coherent access to their critical regions?  It seemed that things could get very confusing and chaotic.  
The way out that we found in this work was to use an abstraction---a conceptual \emph{Virtual Supervisor} process, which could be emulated by the other processes.  The role of the Virtual Supervisor was to offload the coordination information into its own state instead of using the small shared variable, and to manage the coordination via communication with the competing processes using the narrow-bandwidth shared variable. 
In this way, we might say that this paper introduced the important idea of levels of abstraction for distributed algorithms.  

The algorithms of~\available{\cite{BurnsJLFP82}} were not particularly practical.  But they used some interesting algorithmic strategies, mainly the Virtual Supervisor idea.
%
%
The main novelty of the paper was in the lower bound results and proofs, for mutual exclusion with different fairness guarantees. These used explicit, intricate constructions of executions that lock out some processes, given too few values of the shared variable.  These results were inspired by, and generalized considerably, the $2$-process construction of Cremers and Hibbard.
This work reinforced the feasibility of proving lower bounds in this area.

We continued our work with many papers delineating the power of the test-and-set shared memory model to solve problems related to mutual exclusion.  For example, in~\available{\cite{4568019}}, we extended the model and problem to consider multiple resources and process failures. 

\paragraph{Limitations for read/write shared memory:}
The next paper, Paper 3.2~\available{\cite{DBLP:journals/iandc/BurnsL93}}, deviated from the test-and-set model, assuming instead much weaker \emph{read/write} shared memory, in which variables are accessed using only separate read and write operations. The main difficulty with this model is that processes can overwrite each other's updates, which can cause information to be lost.  We again studied the mutual exclusion problem, this time without any fairness requirement, just absence of deadlock.
For the read/write case, unlike the test-and-set case, it turns out the number of variables matters.
So now we focused on the number of variables needed, rather than their size.

The results of~\available{\cite{DBLP:journals/iandc/BurnsL93}} originally appeared in Jim Burns's Georgia Tech PhD thesis, around 1981.  However, for some reason, we did not get around to publishing this until a full $12$ years later. \\

\thepapers{
\noindent
3.2.  James E. Burns and Nancy A. Lynch. Bounds on shared memory for mutual exclusion. Information and Computation, 107(2):171--184, December 1993.\\
}

The paper contains a simple $n$-process algorithm using only $n$ single-writer, multi-reader, binary valued shared variables.  This is in contrast to Lamport's Bakery Algorithm, which has unbounded-size variables, and to Dijkstra's mutual exclusion algorithm, which uses multi-writer shared variables.

The more interesting result in the paper (to me) is the lower bound.  It says that $n$ processes require at least $n$ shared read/write variables to solve this version of mutual exclusion.  No matter how many values the variables may take on, and even if the variables allow multiple writers, we still need at least as many variables as processes!

The proof of this lower bound is by a really clever construction, mainly due to Jim. The key is that we can maneuver a process so that it is poised, "about to write" a shared variable.  
Then we can force activity of the other processes, affecting that variable, then resume the poised process and overwrite whatever was written by the other processes.  This allows us to hide the activity of the other processes.
If we can maneuver multiple processes so they are poised, "about to write" several different shared variables, then we can hide the activity of other processes that affect all of those variables.

\subsection{Models for distributed systems}

Even in our earliest days of working on algorithms and lower bounds for shared-memory distributed systems, we knew we needed new formal models to support the algorithmic work. 
We were already aware that we wanted to help start a new research field of distributed computing theory, and it seemed clear that the field should have its own general models to provide a foundation.

The field of sequential complexity theory was based on well-established automata-theoretic models, like Turing machines and Random Access Machines (RAMs).  Researchers working on sequential algorithms shared a common foundation for their algorithmic work.  For synchronous parallel shared-memory computing, researchers used models such as Parallel Random Access Machines (PRAMs).  
But now we needed something quite different---a model for algorithms that were supposed to run on distributed systems consisting of asynchronously operating, interacting processes.  It was not obvious at all what such a model should look like. 

We did know that we wanted to base our model on a foundation of set theory and automata theory, rather than on any particular logical language.
This was consistent with the situation for sequential algorithms and synchronous parallel algorithms.
But it contrasted with a large body of work that was going on at the time on models for concurrent systems, by Hoare, Milner, and other \emph{process algebraicists}. 
Their work described concurrent processes using formal logical expressions, rather than using automata.
Systems could be built up from simpler components using formally-defined operators, composition being the most important one.
The operators became part of an algebraic language for describing systems of processes, and algebraic equations were used to assert equivalence of systems built using different algebraic expressions.  
Essentially all reasoning about the systems was carried out within a formal logical system. 

But since we were emphasizing algorithms and complexity, we preferred a different style, based on set theory and automata.  We did not want to force the reasoning about the algorithms---their correctness and performance---into any particular logical framework, but rather, wanted to be able to use the full power of mathematics.   

\paragraph{General model for asynchronous shared-memory computing:}
Since we started working on theory for distributed computing by studying shared-memory algorithms, we defined our first model for systems of asynchronous processes interacting using shared memory.  Our initial modeling paper was Paper 3.3~\available{\cite{DBLP:journals/tcs/LynchF81}}: \\

\thepapers{
\noindent
3.3.  Nancy A. Lynch and Michael J. Fischer. On describing the behavior and implementation of distributed systems. Theoretical Computer Science, 13(1):17--43, 1981. Special issue on Semantics of Concurrent Computation.} \\

The paper is written as a sort of manifesto about what a theory for distributed systems should look like.  I think that our viewpoint in this paper turned out to be a reasonably good predictor of how the modeling part of the field would develop.  The paper outlines our goals in terms of creating a general model as the foundation for a new algorithmic theory for distributed systems, and explains our particular design choices. 
It explains why we need new models for distributed algorithms, different from those for sequential algorithms.  We also need new kinds of problem definitions---not just functions as before, but definitions that include ongoing behavior and nondeterminism.  We also need new definitions of what it means for an instance of the model to "solve" a particular problem---something that is obvious for sequential algorithms.


Section 2 of the paper contains the formal definitions for our automata-theoretic shared-memory model.  It also introduces a \emph{composition operator} for systems, which is intended to be useful in describing systems in terms of more primitive systems.  Section 3 describes how to define a \emph{problem} to be solved by an automaton, as a set of (finite and/or infinite) sequences of steps involving shared variables. 
Section 4 describes what it means for a distributed algorithm, expressed as an automaton within our model, to \emph{solve} a problem, also expressed within our model. 
Specifically, the set of sequences comprising the behavior of the algorithm can be any subset of the set of sequences defining the problem. 

The paper continues by defining a useful measure of \emph{time complexity} for asynchronous distributed systems, based on assuming real-time upper bounds on the time between basic events such as processes accessing shared variables.  
%
In contrast to the usual handling for sequential algorithms, we do not consider the costs of local computation, which are generally regarded as insignificant for distributed algorithms compared to the costs of interaction.
The paper goes on to present examples illustrating how very different distributed algorithms can be used to solve the same problem---here, the problem of fair mutual exclusion.  It includes analysis of the time complexity of the three algorithms, and a comparison.

As it turned out, the particular model of this paper did not end up seeing widespread use as a foundation for the field.
However, it was a direct precursor for the more impactful \emph{Input/Output Automata} model of Lynch and Tuttle~\available{\cite{DBLP:conf/podc/LynchT87,LynchT89}}.
Note that the I/O Automata work includes treatment of \emph{levels of abstraction}.
Levels of abstraction have turned out to be another important way of decomposing distributed algorithms, along with composition, but they do not appear in~\available{\cite{DBLP:journals/tcs/LynchF81}}.
I discuss I/O Automata in Section~\ref{sec: earlyMIT-models}.

\paragraph{Synchronous vs. asynchronous shared memory systems:}
While we were working on our algorithms and models for shared-memory distributed computing, Prof. Eshrat Arjomandi from York University visited us at Georgia Tech.  She had been working on parallel algorithms for solving graph problems, using PRAM models.  Like the models we were considering, the PRAM model is a shared-memory parallel model.  However, unlike in our model, its processes operate in synchronous rounds.  See, for example,~\cite{ArjomandiC75}.

This led us to explore the difference in computing power, specifically, in time efficiency, between  synchronous shared-memory models in which the processes operate in "rounds", and asynchronous shared-memory models in which processes operate at arbitrary speeds relative to each other. 
This eventually led to Paper 3.4~\available{\cite{ArjomandiFL83}}. 
This paper contains the perhaps-surprising result that there are some problems that can be solved much faster with a synchronous parallel shared-memory algorithm than with \emph{any} asynchronous shared-memory algorithm. This is not a comparison of the behavior of a particular synchronous algorithm with a particular asynchronous algorithm---rather, it is a result about \emph{all possible} asynchronous algorithms in our model.  \\

\thepapers{
\noindent
3.4.  Eshrat Arjomandi, Michael J. Fischer, and Nancy A. Lynch. Efficiency of synchronous versus asynchronous distributed systems. Journal of the ACM, 30(3):449--456, July 1983. \\
}

The problem we focused on was a simple abstract problem that we called the "$s$-session problem".  In this problem, all of the processes should cooperate to perform a number $s$ of "sessions", in each of which each process must output at least one signal.  Then all processes must halt, producing no further outputs. 
It is easy to devise a synchronous algorithm that performs $s$ sessions in $s$ rounds, i.e., in time $s$.  However, we show that any asynchronous algorithm that is guaranteed to produce $s$ sessions and then halt must take at least approximately $s \log{n}$ time, where $n$ is the number of processes.  Here time is measured according to the asynchronous time measure described in Paper 3.3~\available{\cite{DBLP:journals/tcs/LynchF81}}. 

To prove the lower bound for asynchronous algorithms, we assumed an asynchronous algorithm $\cal A$ that worked in less time than what we wanted to show.  We started with a synchronous execution of $\cal A$.  Then we reordered the steps of the execution, while maintaining the dependencies of the algorithm and reducing the number of sessions to below $s$. 
The reordered execution is still an execution of the algorithm, but it fails to solve the $s$-session problem.  This yielded a contradiction.

The significance of this paper was that it removes the following hope:  we might like to design asynchronous distributed algorithms by first designing synchronous versions and then transforming them systematically to run in asynchronous systems, while preserving correctness.  This result says that such an approach would increase the worst-case running time significantly, at least for some problems.

\subsection{Other early results}

\paragraph{Resource allocation in networks:}
With Nancy Griffeth, we studied more elaborate resource-allocation problems~\available{\cite{DBLP:journals/iandc/LynchGFG86, DBLP:journals/iandc/FischerGGL92}}.
We assumed an arbitrary placement of resources at the nodes of a graph network. The problem was to service requests that arrive at random nodes in the network, matching requests to resources.
The model used in this work was a network message-passing model rather than a shared-memory model. 

The first paper~\available{\cite{DBLP:journals/iandc/LynchGFG86}} presented request-resource matching algorithms and analyzed their time requirements.
We assumed here that requests arrive on-line, and may overlap:  new requests may arrive before previous requests have been matched to their resources.
The worst case for this algorithm turned out to be when the requests arrive sequentially, because some optimization is possible when concurrent requests interact.
The second paper~\available{\cite{DBLP:journals/iandc/FischerGGL92}} assumed that the requests arrive at random locations, all at once.
Here we measured the sum of the lengths of paths connecting requests to their matched resources.

\paragraph{Global states:}
We also wrote a paper describing a new \emph{global snapshot} algorithm for distributed systems\available{~\cite{DBLP:journals/tse/FischerGL82}}.
This paper was inspired by work on distributed database concurrency control.  
We assumed a system of "ordinary transactions", which could execute in an interleaved manner, and added to this system a new "snapshot transaction", which was required to appear atomic with respect to the ordinary transactions.  The snapshot transaction was required to return a "consistent" global state of the system---one that "could have happened" as a result of running all the ordinary transactions that preceded the snapshot to completion, along with some subset of the ordinary transactions that were concurrent with the snapshot.  
In this sense, the algorithm produced a \emph{consistent global snapshot}.

A special case of this snapshot algorithm is one in which the "ordinary transactions" correspond to transfers of money from one banking location to another.  The snapshot transaction can then be used to calculate the total money at all of the banking locations.

\subsection{Distributed consensus}

Much of our best-known early work was on the topic of distributed consensus.  
In this problem, processes receive inputs, which may differ, and must agree on a common output.
This must work even if some number (at most $f$) of processes are faulty.
In some versions of the problem, they might just stop.
Or they might exhibit worse, \emph{Byzantine} failures, where they act "maliciously", for example, providing inconsistent information to different other processes.
This research direction was initiated by earlier work of Leslie Lamport with Marshall Pease and Robert Shostak~\cite{DBLP:journals/jacm/PeaseSL80, DBLP:journals/toplas/LamportSP82}, on agreement for altimeter readings on aircraft, where one of the altimeters might fail.

Leslie visited Mike Fischer and myself at Georgia Tech, for about a week during Mike's sabbatical.
In preparation for Leslie's visit, Mike and I read a number of Leslie's papers, including a preliminary (unpublished) manuscript entitled "The Albanian Generals Problem" (the problem was later renamed to the "Byzantine Generals Problem").  
The paper considered a synchronous model in which the processes acted in rounds.  Communication was by sending and receiving messages.
Processes were totally connected for communication.

In reading these papers, we noted that all of the algorithms in the papers used exactly $f+1$ rounds to reach agreement tolerating $f$ faulty processes.   
We wondered whether this many rounds were necessary.
We tried to find faster algorithms, but failed.
Then just before Lamport's visit, we proved a lower bound of $f+1$ rounds.  
This result appears in the short Paper 3.5~\available{\cite{DBLP:journals/ipl/FischerL82}} \\

\thepapers{
\noindent
3.5.  Michael J. Fischer and Nancy A. Lynch. A lower bound for the time to assure interactive consistency. Information Processing Letters, 14(4):183--186, June 1982.} \\

The interesting idea of this proof was what later became known as \emph{chain argument}.  We constructed a chain of executions spanning between two extreme executions in which the decisions for the nonfaulty processes are required to be $0$ and $1$, respectively.  
In this case, we considered an execution with all inputs of $0$ and no failures, vs. an execution with all inputs of $1$ and no failures.  
Then we constructed the chain, one "link" at a time, so that at each link between two executions, there is some nonfaulty process that cannot distinguish between the two executions.
This implies that this process, and hence all the nonfaulty processes, must decide in the same way in both executions.
It follows that there can be no particular point in the chain where the decision changes, which yields the contradiction.

Another version of this proof appears in my Distributed Algorithms textbook~\available{\cite{DBLP:books/mk/Lynch96}};
That version proves a stronger result, for stopping failures, not just very strong Byzantine failures. 

Chain arguments were used later to prove many other lower bound results.  One forms the basis of my best-known result, the "FLP" impossibility result~\available{\cite{DBLP:journals/jacm/FischerLP85}}, which I discuss in Section~\ref{subsection: FLP}.
Also, chain arguments have been extended to multiple dimensions, using topological machinery.  
Such arguments appear in the Godel-prize winning work on the topological structure of asynchronous computability~\cite{DBLP:journals/jacm/HerlihyS99}, as well as our lower bound for synchronous solutions to $k$-consensus~\available{\cite{DBLP:conf/focs/ChaudhuriHLT93}}, which I discuss in Section~\ref{sec: laterMITalgorithms}.

Thus, Paper 3.5~\available{\cite{DBLP:journals/ipl/FischerL82}} marked the beginning of our work on consensus, a topic that we continued to study for quite a few years.\\
\\

In these early papers, we have already seen several lower bounds and other impossibility results.  Impossibility results have turned out to be a major research direction within distributed computing theory, with hundreds of such results being proved.  
Such impossibility results depend on the difficulty of the distributed platforms on which the algorithms run, in particular, on the strong limitations of locality in distributed algorithms.
Each process can see only its own state, the values it reads in shared memory, the messages it sends and receives, etc. The lack of global knowledge is a very strong restriction that enables impossibility results.  
We will see more lower bounds in this paper, and others are discussed in the papers~\available{\cite{10.1145/72981.72982}} and~\cite{ER2003}.

We did other projects in distributed computing theory at Georgia Tech, but the ones I have described should give you the general flavor.
After getting tenure at Georgia Tech, I took a sabbatical at MIT from 1981-82, continuing to work on developing the new field of distributed computing theory.  I will discuss this next.

\section{Early Work at MIT:  Distributed Consensus, Models, and Atomic Transactions, 1981-1990}
\label{sec: earlyMIT}

I went to MIT for a sabbatical year in 1981-1982, visiting Prof. Barbara Liskov's research group in distributed systems.  
I was interested in distributed database transaction management algorithms;
Barbara and her group were working on a system called Argus, which implemented a programming model based on nested atomic transactions.
A transaction is a sequence of operations on data object that are supposed to be performed so that they look "as if" they occurred atomically, that is, consecutively; however, in the implementations, the operations are generally interleaved.
Nested transactions generalize the traditional transaction model by allowing transactions to have subtransactions, and so on.
Many interesting algorithms had been developed in the distributed systems community for implementing transactions in distributed systems, including algorithms based on locking, time-stamping, and hybrids of these, optimistic algorithms, and algorithms that used data replication.

I was interested in working on theory for such systems.
I worked with Barbara, and with her group members Bill Weihl and Maurice Herlihy, to model and verify some popular algorithms for distributed database concurrency control, including algorithms based on locking, timestamping, optimistic concurrency control, and replication. 
This work involved generalizing the standard concurrency control algorithms to the case where transactions could be nested; Barbara had already done that for locking, but there was more to do for other algorithms.  In addition to algorithm design, this turned into an exercise in formal modeling and verification of complex distributed algorithms with strong correctness requirements.  

In addition, I continued work on distributed consensus and consensus-related problems, mainly with Mike Fischer.  This work culminated at the end of the year with the famous "FLP" result on impossibility of consensus. 

Another interesting event during this sabbatical year was the start of the Principles of Distributed Computing (PODC) conference.   The first of these meetings was held in Ottawa in the summer of 1982, and the conference has been going strong since then.  As I recall, I played some role in starting up this conference, while the main organizers were Mike Fischer, Robert Probert, and Nicola Santoro.

At the end of the sabbatical year, I received and accepted an offer of a tenured Associate Professorship at MIT, where I have remained for the 40+ years since then.
Along with the offer, I was given a five-year chair, the first "Ellen Swallow Richards Chair", funded by the MIT Alumnae Association with the purpose of bringing senior women faculty to MIT.  At that time, there were very few. 

For the next few years, I built my research group and worked on a combination of consensus algorithms and lower bounds, and nested transaction algorithms and system modeling.  The work on nested transactions contributed, in turn, to the development of the Input/Output automaton model for asynchronous systems communicating via shared actions.  

Although this was all theoretical work, I remained for these years in the Distributed Systems group at MIT, since distributed computing was still not a popular theoretical discipline.

\subsection{Key publications}

\thepapers{
4.1.  Michael J. Fischer, Nancy A. Lynch, and Michael S. Paterson. Impossibility of distributed consensus with one faulty process. Journal of the ACM, 32(2):374--382, April 1985. \\

\noindent
4.2.  Danny Dolev, Nancy A. Lynch, Shlomit S. Pinter, Eugene W. Stark, and William E. Weihl. Reaching approximate agreement in the presence of faults. Journal of the ACM, 33(3):499--516, July 1986. \\

\noindent
4.3.  Jennifer Lundelius and Nancy Lynch. An upper and lower bound for clock synchronization. Information and Control, 62(2-3):190--204, August/September 1984.\\

\noindent
4.4.  Michael J. Fischer, Nancy A. Lynch, and Michael Merritt. Easy impossibility proofs for distributed consensus problems. Distributed Computing, 1(1):26--39, January 1986. \\

\noindent
4.5.  Cynthia Dwork, Nancy Lynch, and Larry Stockmeyer. Consensus in the presence of partial synchrony. Journal of the ACM, 35(2):288--323, April 1988. \\

\noindent
4.6. Nancy A. Lynch and Mark R. Tuttle. Hierarchical correctness proofs for distributed algorithms. In Proceedings of the Sixth Annual ACM Symposium on Principles of Distributed Computing, pages 137--151, Vancouver, British Columbia, Canada, August 1987.] \\

\noindent
4.7.  Nancy Lynch and Mark Tuttle. An introduction to Input/Output Automata. CWI-Quarterly, 2(3):219--246, September 1989. Centrum voor Wiskunde en Informatica, Amsterdam, The Netherlands. \\

\noindent
4.8.  Nancy A. Lynch, Michael Merritt, William E. Weihl, and Alan D. Fekete.
Atomic Transactions. Morgan Kaufmann series in data management systems. Morgan Kaufmann, 1993.
}

\subsection{FLP}
\label{subsection: FLP}

Paper 4.1~\available{\cite{DBLP:journals/jacm/FischerLP85}} is, by far, my best known and  most cited paper, with 7015 citations at this moment.  
The main result, and the paper, are generally known as FLP, after the authors. 
We obtained the result in late summer of 1982, at the end of my sabbatical year and just before I started my faculty job at MIT.  We first published the result in a Principles of Database Systems conference in 1983~\available{\cite{DBLP:conf/pods/FischerLP83}}.  The final journal version appeared in JACM in 1985~\available{\cite{DBLP:journals/jacm/FischerLP85}}. 

The result says that it is impossible to reliably reach agreement in an asynchronous network, with the possibility of even a single, simple processor stopping failure.\\

\thepapers{
\noindent
4.1.  Michael J. Fischer, Nancy A. Lynch, and Michael S. Paterson. Impossibility of distributed consensus with one faulty process. Journal of the ACM, 32(2):374--382, April 1985. \\
}

As I recall, I thought of this, at first, as a purely theoretical problem. 
Having matching upper and lower bounds for the number of rounds needed for consensus in synchronous distributed systems~\available{\cite{DBLP:journals/ipl/FischerL82}}, it seemed natural to consider what happens in the asynchronous case. 
I did not know how this would turn out ahead of time; I worked both on trying to find an algorithm and trying to prove an impossibility result.  Some sense that this might be solvable arose from our early thoughts on the \emph{approximate agreement problem}, discussed in Section~\ref{sec: more consensus}, for which natural synchronous algorithms had straightforward extensions to the asynchronous setting.

Mike Fischer and I met at MIT during the summer of 1982 and discussed this problem.  Mike said that he had also been thinking about the same problem, in his case motivated by the need for such an algorithm in certain distributed database systems.  Apparently, Butler Lampson had suggested this problem to Mike during a visit by Mike to Xerox PARC.
Mike Paterson joined the collaboration when he visited Mike Fischer at Yale sometime after our MIT meeting.
We worked both on trying to find an algorithm and on trying to prove an impossibility result, and ended up proving our impossibility result.

The FLP result was surprising, since it required only one failure, of a very simple kind.
The proof was elementary, but not at all obvious. 
In fact, as I recall, some leading researchers couldn't quite believe it was true and had to review it several times.
The proof was based on assuming that a solution exists, and then characterizing how the decision could be made in such a solution.  It turns out that the decision can be localized to what happens at a single location in the network, based simply on the order of arrival of two different messages at one node.  But if the node in question fails, the rest of the system cannot distinguish the order in which the messages arrived, so cannot decide differently in the two cases.

As it turned out, the result was widely appreciated in both the distributed computing theory and distributed systems communities.
Certainly it was an interesting, surprising theoretical result.  
But also, it had significance for the practical distributed systems community.  Namely, the problem was closely related to the distributed transaction commit problem, where the processes involved in processing a transaction must agree on whether the transaction should commit or abort.
Systems researchers were trying to design algorithms to solve the commit problem, and were arguing informally that their algorithms were correct.  Our impossibility result seemed to imply that they must have been wrong, since the problem was fundamentally unsolvable.

I admit to being surprised by how much attention this result got in the distributed systems community.  Though I was quite pleased with the result and proof as theory.

Our paper was awarded the second Dijkstra Prize (after Lamport's famous "Time, Clocks,..." paper).
Here is a citation for that award, written by Jennifer Welch and Nir Shavit:

\begin{quote}
The result of this paper (commonly known as FLP) is that, surprisingly, it is impossible for a set of processors in an asynchronous distributed system to agree on a binary value, even if only a single processor is subject to an unannounced crash. Although the result was motivated by the problem of committing transactions in distributed database systems, the proof is sufficiently general that it directly implies the impossibility of a number of related problems, including consensus.

This result has had a monumental impact in distributed computing, both theory and practice. Systems designers were motivated to clarify their claims concerning under what circumstances the systems work.

On the theory side, people have attempted to get around the impossibility result by changing the system assumptions or the problem statement. Work on changing the system assumptions includes the study of partially synchronous models and of various kinds of failure detectors. Modified problem statements include randomized algorithms, approximate agreement, $k$-set agreement, and condition-based approaches.

The proof technique used in FLP, valency arguments, has been used and adapted to show many other impossibility and lower bound results in distributed computing. These include impossibility results for consensus, $k$-set consensus, and renaming in various models, and lower bounds on contention and on the number of rounds for synchronous consensus.

The FLP result forms the basis of work on the wait-free hierarchy, in which data types are classified and compared according to the maximum number of processes for which they can solve wait-free consensus. The calculation of consensus numbers relies on valency arguments.

Finally, work on applying ideas from topology to fault-tolerant distributed computing were inspired by the posing of the $k$-set consensus problem, which in turn was inspired by the FLP result.
\end{quote}

\subsection{More results related to consensus}
\label{sec: more consensus}

During the sabbatical year 1981-1982 and the next few years at MIT, we continued working on problems related to distributed consensus.  
There were many questions to consider, and this general topic became a popular direction for theoretical computer science research during that time.
My collaborators during this time were my own group members, plus others including Mike Fischer, Bill Weihl, and Jim Burns.

\paragraph{Low-communication consensus:}
One of our first results involved improving the amount of communication needed for consensus in synchronous systems.  The papers by Lamport et al.~\cite{DBLP:journals/jacm/PeaseSL80, DBLP:journals/toplas/LamportSP82} gave algorithms for Byzantine agreement that used an exponential amount of communication---the number of bits exchanged was exponential in the number of faulty processes that were tolerated.  
Dolev and Strong~\cite{10.1145/800070.802215} obtained a polynomial-communication algorithm, and we improved this further, obtaining an algorithm that required only $O(nt + t^3 \log{t})$ bits\available{~\cite{DBLP:journals/iandc/DolevFFLS82}}.

\paragraph{Approximate agreement:} 
Another consensus-related problem that we considered was that of \emph{approximate agreement}, which we introduced in Paper 4.2 ~\available{\cite{DBLP:journals/jacm/DolevLPSW86}}.  \\

\thepapers{
\noindent
4.2.  Danny Dolev, Nancy A. Lynch, Shlomit S. Pinter, Eugene W. Stark, and William E. Weihl. Reaching approximate agreement in the presence of faults. Journal of the ACM, 33(3):499--516, July 1986.\\}

This problem is a variant of the Byzantine Generals problem in which processes start with arbitrary real values rather than values from a discrete domain, and in which they must agree approximately rather than exactly.  We devised algorithms to reach approximate agreement in both synchronous and asynchronous systems.
The asynchronous agreement algorithm is especially interesting because it contrasts with the FLP impossibility result, which says that exact agreement with guaranteed termination is not attainable in an asynchronous system with even one faulty process.  
It turns out that the situation is quite different if we require only approximate agreement.

The algorithms work in rounds; for the asynchronous case, this required defining a new fault-tolerant notion of asynchronous rounds. 
The algorithms work by successive approximation.
At each round, they use a new \emph{fault-tolerant averaging function}---one that discards a number of extreme values corresponding to the number of process failures to be tolerated. 
We proved a convergence rate that depends on the ratio between the number $f$ of faulty processes and the total number $n$ of processes.   We also proved lower bounds on the convergence rate, which imply that our algorithms are optimal.

\paragraph{Clock synchronization:}
We also studied the problem of \emph{distributed clock synchronization}.  
This used a model that is somewhere between synchronous and asynchronous.
Namely, processes have individual real-valued "clocks" that they can read and use in determining their behavior. 
The clocks of different processes may differ slightly from each other, and they may run at rates that differ slightly from that of real time.  Therefore, they may drift apart.  Other complications include (possibly Byzantine) faulty processes and variations in communication time.
It is desirable to keep the clocks synchronized, as closely as possible, by making small adjustments or using them to implement approximately-synchronized \emph{logical clocks}.   

As I recall, the approximate agreement problem that we introduced in~\available{\cite{DBLP:journals/jacm/DolevLPSW86}} was originally inspired by the clock synchronization problem, since it may be viewed (roughly speaking) as a simpler special case.

A collection of fault-tolerant clock synchronization algorithms appeared in~\cite{10.1145/2455.2457, DBLP:conf/podc/MarzulloO83,DBLP:conf/podc/HalpernSSD84}.  My PhD student Jennifer Lundelius (Welch) and I contributed to this direction~\available{\cite{DBLP:journals/iandc/WelchL88}} with a new clock synchronization algorithm that was directly inspired by our asynchronous approximate agreement algorithm of~\available{\cite{DBLP:journals/jacm/DolevLPSW86}}.
As in that algorithm, the new clock synchronization algorithm proceeds in rounds, adjusting the clocks using a fault-tolerant averaging function.

In addition, in Paper 4.3~\available{\cite{DBLP:journals/iandc/LundeliusL84}},
Jennifer and I proved matching upper and lower bound results for a simple special case of the problem, in which clocks do not drift and there are no failures.  However, the clocks at different processes may have different initial values, and the communication delay between the processes is uncertain. 
In terms of the number $n$ of processes and a bound $\epsilon$ on the uncertainty in communication delay, we obtained matching upper and lower bounds of $O(\epsilon(1 - 1/n))$ for the achievable closeness of synchronization. 
The basic technique involves constructing alternative executions by shifting the times of events at different processes.  \\

\thepapers{
\noindent
4.3.  Jennifer Lundelius and Nancy Lynch. An upper and lower bound for clock synchronization. Information and Control, 62(2-3):190--204, August/September 1984.\\
}

In another interesting paper on fault-tolerant clock synchronization around the same time,
Dolev, Halpern, and Strong~\cite{DBLP:journals/jcss/DolevHS86} proved impossibility of Byzantine-fault-tolerant clock synchronization for $3f$ processes and $f$ failures.  We noted that the Byzantine agreement algorithms of Lamport et al. also used $3f+1$ processes, and the authors proved that this was necessary.
These results suggest that there might be something inherent about needing $3f + 1$ processes to tolerate $f$ Byzantine failures, in general. 
I will have more to say about this below.

\paragraph{The Byzantine Firing Squad problem:}
Jim Burns and I invented a new synchronization problem, called the \emph{Byzantine Firing Squad} problem~\available{\cite{preparata1987advances<}}, and developed a new algorithm to solve it.  
The problem assumes that processes operate in synchronous rounds but do not have a common start time.  We assumed that one or more nonfaulty processes receive an external START signal at some point, and all the nonfaulty processes are supposed to fire at some later round.  Moreover, if any nonfaulty process fires in some round, then all the nonfaulty processes must fire in that same round.
As for the previous Byzantine agreement and Byzantine clock synchronization algorithms, our Byzantine Firing Squad algorithm also uses $3f+1$ processes to tolerate $f$ Byzantine failures. 
This was not surprising, because our algorithm uses a standard Byzantine agreement algorithm to agree on a firing preference at every round.

\paragraph{Easy impossibility proofs:}
\label{sec: more-imposs}

We (and others) observed the common $3f+1$ processes bound for Byzantine fault-tolerant algorithms, including Byzantine agreement, Byzantine clock synchronization, and the Byzantine Firing Squad problem. 
All the known algorithms for these problems used $3f + 1$ processes to tolerate $f$ faults.
Earlier papers~\cite{DBLP:journals/toplas/LamportSP82} and~\cite{DBLP:journals/jcss/DolevHS86} included proofs that $3f+1$ is a lower bound, for Byzantine agreement and clock synchronization, respectively.  These were special-case proofs, for particular problems.

Mike Fischer, Michael Merritt, and I thought that there must be some common reason that $3f+1$ was necessary for accomplishing anything interesting with Byzantine faults.
In Paper 4.4~\available{\cite{DBLP:journals/dc/FischerLM86}}, we described a systematic way to prove such lower bounds, and applied it to five problems:  Byzantine agreement, weak Byzantine agreement~\cite{10.1145/2402.322398}, approximate agreement, clock synchronization, and the Byzantine Firing Squad problem. I think that our approach yielded some unifying insight. \\

\thepapers{
\noindent
4.4.  Michael J. Fischer, Nancy A. Lynch, and Michael Merritt. Easy impossibility proofs for distributed consensus problems. Distributed Computing, 1(1):26--39, January 1986. \\
}

The key idea of the approach can be expressed nicely at a high level.
Consider the special case of showing that three processes cannot solve Byzantine agreement if one of the processes might be faulty.  Assume for contradiction that such a system of three processes, $\mathcal A$, exists.  Now construct another system $\mathcal S$, this one of six processes, which is essentially two copies of the 3-process system $\mathcal A$.  
In $\mathcal S$, start half the processes with input $0$ and the other half with input $1$. 
Then, under some reasonable assumptions, it turns out that we can deduce behaviors for the $6$-process system $\mathcal S$ by using the correctness requirements of the $3$-process system $\mathcal A$.  This depends on the fact that processes in $\mathcal S$ cannot distinguish themselves from their counterparts in $\mathcal A$, provided that a Byzantine faulty process in $\mathcal A$ pretends to be a string of four processes in $\mathcal S$.  
Deducing enough behaviors for the $6$-process system $\mathcal S$ yields a contradiction. 

As usual, the proof depends on the limitations of local knowledge in a distributed system.
The proof is expressed generally and abstractly, not dependent on precise details of the model.

Very similar proofs hold for weak Byzantine agreement, approximate agreement, Byzantine clock synchronization, and Byzantine Firing Squad.
In fact, this paper essentially provides a general approach to proving such lower bounds.
In addition to all of these results about the number of required processes, the paper also shows general limitations of $2f+1$ on network connectivity. \\
\\
By 1986, distributed computing theory had become an established field of theoretical study, and lower bounds and other impossibility results had become a main characteristic of the new field.
At some point there were so many, that I felt compelled to write a paper summarizing all of those that I could find at the time~\available{\cite{10.1145/72981.72982}.}  

\subsection{Consensus with partial synchrony} 

The strong impossibility result in the FLP paper was worrisome, because the fault-tolerant consensus problem is very important to solve in practice.
We (and many others) considered how to get around the problem.  Our solution appears in Paper 4.5~\available{\cite{DBLP:journals/jacm/DworkLS88}}.  It involves using notions of \emph{partial synchrony}, which are between pure synchrony and pure asynchrony. \\

\thepapers{
\noindent
4.5.  Cynthia Dwork, Nancy Lynch, and Larry Stockmeyer. Consensus in the presence of partial synchrony. Journal of the ACM, 35(2):288--323, April 1988. \\
}

Our paper covers several types of failures and notions of partial synchrony.  The simplest case involves simple stopping failures and a notion of partial synchrony in which message delays are bounded after some \emph{Global Stabilization Time (GST)}.
The algorithm for this case requires a majority of nonfaulty processes.

The first key idea in the paper was to give the safety requirements (agreement and validity) higher priority than the termination requirement.  The safety requirements should never be violated, regardless of the occurrence of failures or timing anomalies.  However, the termination requirement might be permitted to depend on stability in the system's behavior, with respect to timing and failures. 
This seemed reasonable, for use in practice.

The second key idea was to have the algorithm make multiple attempts to achieve consensus, where each attempt involved a protocol, led by a coordinator, to gather enough votes for agreement on a particular value.  If the system stabilized for long enough, then an attempt initiated during the stable period would be guaranteed to complete.

The danger of this approach is that multiple attempts could lead to contradictory decisions by different processes.  This motivated the third key idea, which was a mechanism designed to keep the results of different partial attempts consistent.
For this, we used a protocol in which processes can "lock" proposed values. Each lock is associated with a particular consensus attempt.  A process can lock a value for a certain attempt when it learns that the coordinator of that attempt has proposed that value for a decision value.  A process can release a lock when it hears about a lock for a different value for a later attempt.

To determine a value to propose, the coordinator of an attempt gathers information from a majority of processes.  The requirement is that each of the processes in the majority must deem the value to be "acceptable", meaning that it doesn't have a lock on any different value.
Once the coordinator determines a value to propose, it sends messages to all processes, requesting that they lock this value for this attempt.
If it hears that a majority have done this, the coordinator can decide on this value.
This protocols requires a majority of nonfaulty processes.

This paper was a precursor of the well-known Paxos algorithm~\cite{10.1145/279227.279229}.
Paxos used the same basic idea for consensus, but incorporated this consensus protocol into a larger protocol for implementing an ongoing replicated state machine; basically, Paxos achieves consensus on each successive update. 
Also, Paxos was designed to tolerate more concurrency in the attempts---our attempts were sequential---but the same sort of consistency mechanism still works.
Paxos is also designed to tolerate more practical types of failures, such as crashes that obliterate volatile memory.  Variants of Paxos have been engineered and widely used in practice.  
The Paxos algorithm was, in turn, a precursor of modern blockchain algorithms.

The significance of the work in~\available{\cite{DBLP:journals/jacm/DworkLS88}} was well recognized in both the theoretical and practical distributed systems communities.
Although we designed this as a theoretical result, it turned out to have considerable significance in practice.  This paper was awarded the 2007 Dijkstra Prize.
Here is the citation (which I think was written by Dahlia Malkhi):

\begin{quote}
This paper introduces a number of practically motivated partial synchrony models that lie between the completely synchronous and the completely asynchronous models, and in which consensus is solvable. It gave practitioners the right tool for building fault tolerant systems, and contributed to the understanding that safety can be maintained at all times, despite the impossibility of consensus and progress is facilitated during periods of stability. These are the pillars on which every fault tolerant system has been built for two decades. This includes academic projects such as Petal, Frangipani, and Boxwood, as well as real life data centers, such as the Google file system.

In distributed systems, balancing the pragmatics of building software that works against the need for rigor is particularly difficult because of impossibility results such as the FLP theorem. The publication by Dwork, Lynch, and Stockmeyer was in many respects the first to suggest a path through this thicket, and has been enormously influential. It presents consensus algorithms for a number of partial synchrony models with different timing requirements and failure assumptions: crash, authenticated Byzantine, and Byzantine failures. It also proves tight lower bounds on the resilience of such algorithms.

The eventual synchrony approach introduced in this paper is used to model algorithms that provide safety at all times, even in completely asynchronous runs, and guarantee liveness once the system stabilizes. This has since been established as the leading approach for circumventing the FLP impossibility result and solving asynchronous consensus, atomic broadcast, and state-machine replication.

In particular, the distributed systems engineering community has been increasingly drawn towards systems architectures that reflect the basic split between safety and liveness cited above. Dwork, Lynch, and Stockmeyer thus planted the seed for a profound rethinking of the ways that we should build, and reason about, this class of systems. Following this direction are many foundational solutions. First, these include state machine replication methods such as Lamport’s seminal Paxos algorithm and many group communication methods. Another important branch of research that directly follows this work is given by Chandra and Toueg’s unreliable failure detector abstraction, which is realized in the eventual synchrony model of this paper. As Chandra and Toueg write: “we argue that partial synchrony assumptions can be encapsulated in the unreliability of failure detectors. For example, in the models of partial synchrony considered in Dwork et al. it is easy to implement a failure detector that satisfies the properties of Diamond-W.” Finally, the insight by Dwork, Lynch, and Stockmeyer also led to various timed-based models of partial synchrony, such as Cristian and Fetzer’s Timed-Asynchronous model and others. 
\end{quote}

\subsection{Models}
\label{sec: earlyMIT-models}

In parallel with working on consensus problems, I did a deep dive into the theory of distributed database algorithms, based on \emph{atomic transactions}.  As I described at the beginning of this section, Barbara Liskov introduced me to nested transactions, a generalization of the traditional transaction model that allows transactions to have subtransactions which were atomic with respect to each other, and so on.  Many interesting algorithms were developed for implementing distributed transactions, and all could be extended to the nested case.

I will return to nested transactions once again in Section~\ref{subsection: nested}.  Here, I mention them just as one of my motivations for developing new formal models for asynchronous distributed systems.
Other motivating examples included projects on modeling resource allocation in networks~\available{\cite{DBLP:conf/podc/LynchT87}}, synchronizers~\available{\cite{DBLP:conf/wdag/FeketeLS87}}, distributed minimum spanning tree algorithms~\available{\cite{DBLP:conf/podc/WelchLL88}}, communication channels~\available{\cite{DBLP:conf/podc/LynchMF88}}, shared atomic objects~\available{\cite{DBLP:journals/tc/Bloom88}}, and dataflow systems~\available{\cite{DBLP:journals/iandc/LynchS89}}.

What we needed for all of this work was a new model for asynchronous distributed systems, in which processes communicate with each other, not using shared memory as in~\available{\cite{DBLP:journals/tcs/LynchF81}}, but using input and output actions.  A shared action model seemed better than a shared memory model for network-style examples like those listed above. 
The result was the Input/Output Automata mathematical framework.
We presented this in two papers, Papers 4.6~\available{\cite{DBLP:conf/podc/LynchT87}} and 4.7~\available{\cite{LynchT89}}. \\

\thepapers{
\noindent
4.6. Nancy A. Lynch and Mark R. Tuttle. Hierarchical correctness proofs for distributed algorithms. In Proceedings of the Sixth Annual ACM Symposium on Principles of Distributed Computing, pages 137--151, Vancouver, British Columbia, Canada, August 1987. \\
\\
4.7.  Nancy Lynch and Mark Tuttle. An introduction to Input/Output Automata. CWI-Quarterly, 2(3):219--246, September 1989. Centrum voor Wiskunde en Informatica, Amsterdam, The Netherlands. \\
}

The papers defined I/O Automata, including a treatment of \emph{fair execution} that capture the idea that separate automata, or even \emph{tasks} within automata, continue taking steps independently. 
This notion is essential for a treatment of concurrently executing processes, but did not appear in previous models for concurrent systems.

I/O Automata include separate \emph{input and output actions}.  I/O Automata are \emph{input-enabled}, which means that any input action may occur from any automaton state, in other words, an automaton cannot block its inputs.  
In contrast, in the process-algebraic models that were popular at the time, actions are not classified as input vs. output, which meant that actions that are shared by different automata could cause low-level synchronization delays.
Such delays would not be convenient in a model that is intended to be used for analyzing the time complexity of algorithms.
The distinction between input and output actions, together with the input-enabling property, remove the problem of synchronization delays.

The papers~\available{\cite{DBLP:conf/podc/LynchT87}} and~\available{\cite{LynchT89}} also define \emph{problems} to be solved by I/O Automata; a problem is modeled as a set of sequences of input and output actions.
The papers also define what it means for an I/O automaton to \emph{solve} a problem:  an automaton solves a problem if the set of its external behaviors is a subset of the set defined by the problem.  
The papers also define \emph{composition} for I/O Automata and showed that it respects external behavior.

We also developed machinery for using \emph{levels of abstraction} for describing and proving correctness of distributed algorithms. 
A distributed algorithm might first be described and proved correct in terms of a high-level, abstract model.
Then more detailed, lower-level versions of the algorithm could be defined, and mapped to the abstract model, in a way that preserves externally-visible behavior.
This allows algorithm correctness to carry over automatically from the high-level algorithm to the lower-level versions.
This idea has turned out to be important as a way of understanding and proving correctness of distributed algorithms. 

As an example to illustrate levels of abstraction, in~\available{\cite{DBLP:conf/podc/LynchT87}}, we described a simple "arbiter" algorithm that fairly allocates a single unshareable resource among processes located at the nodes of an undirected acyclic graph. 
We first presented the algorithm at a high level, in terms of request tokens and a resource token moving around the graph. 
This would be the way that someone might explain the operation of the algorithm on the blackboard, except that we gave a formal version. 
Then we gave a lower-level algorithm in terms of actual processes and communication channels. 
We proved that the higher-level algorithm fairly allocates the resource, and that the lower-level algorithm implements the higher-level algorithm, in a formal sense, and so also fairly allocates the resource.

Another paper~\available{\cite{DBLP:journals/iandc/LynchS89}} demonstrates how an important principle studied in some other models of concurrency can be expressed and proved using I/O Automata.  We defined a subclass of I/O Automata called \emph{determinate}, which means just that its input/output relation is a function.  We showed that determinate I/O Automata compute continuous functions; moreover, the function associated with a composition of determinate automata is also continuous, and can be characterized as the least fixed-point of a certain continuous functional associated with the network.  This latter result was known as \emph{Kahn's Principle} in the concurrency theory area. Although the result was already known, our contribution here lies in the fact that the I/O Automata model can express the result easily and yields extremely simple proofs.

The I/O Automata model has been used fairly extensively to describe different types of asynchronous distributed algorithms. 
For example, my two books \emph{Atomic Transactions}~\available{\cite{LynchMWF91}} and \emph{Distributed Algorithms}~\available{\cite{DBLP:books/mk/Lynch96}} use the I/O Automata model as the foundation for presenting all of the asynchronous distributed algorithms contained therein. 
Herlihy used (a slightly simplified version of) I/O Automata as the foundation for his theory for wait-free synchronization~\cite{10.1145/114005.102808}, and Herlihy and Shavit used them for their well-known work on the topological structure of asynchronous computability~\cite{DBLP:journals/jacm/HerlihyS99}.  
Chockler, Keidar and Vitenberg used I/O Automata as a unifying framework to describe and unify many specifications that were developed in the group communications research area~\cite{10.1145/503112.503113}.
Manadhata and Wing used them as the basis for developing metrics for software security~\cite{5482589}.  Abraham et al. used them to explain the Byzantine disk Paxos algorithm~\cite{DBLP:journals/dc/AbrahamCKM06}.  Doherty at al. used them to specify transactional memory~\cite{DBLP:journals/fac/DohertyGLM13}.  And so on.

\subsection{Nested transactions}
\label{subsection: nested}

I have already mentioned our work on nested transactions, more than once.
This was a major effort to understand, in rigorous terms, a collection of algorithms for implementing the nested transactions programming framework for distributed databases. 
My collaborators in this work included Michael Merritt, Bill Weihl, Alan Fekete, Maurice Herlihy, James Aspnes, and my PhD student Ken Goldman.

Our work was originally inspired by Barbara Liskov's Argus system.
Argus used a simple distributed locking algorithm, but we also studied algorithms based on timestamps, algorithms that were hybrids of locking and timestamps, optimistic algorithms, and algorithms that used replicated data. 
Some of these algorithms had been designed for ordinary single-level transactions, but could be extended to allow multiple levels of nesting.

When we began looking at this area, there was little supporting theory; we developed such a theory.
We modeled the requirements of transaction systems, and the many different distributed algorithms that meet these requirements.  We gave complete correctness proofs.  
Overall, we produced a large body of work represented in many papers; I won't go into these results in any detail here.
This work culminated later in the book \emph{Atomic Transactions}~\available{\cite{LynchMWF91}}, Publication 4.8.
The work all rests on the I/O Automata framework.\\

\thepapers{
\noindent
4.8. Nancy A. Lynch, Michael Merritt, William E. Weihl, and Alan D. Fekete.
Atomic Transactions. Morgan Kaufmann series in data management systems.
Morgan Kaufmann, 1993. \\
}

So far, I have described our early work on building a theory for distributed systems, including basic work on theoretical algorithms and impossibility results, formal models for distributed systems, and applications.  This work contributed to establishing distributed computing theory as a research field.

Our main contributions up to 1990 included developing formal foundations for describing distributed algorithms, and using them to describe and prove properties of many theoretical distributed algorithms and practical distributed systems.
We also proved many new impossibility results.  In doing this, we demonstrated that impossibility results are characteristic of the field of distributed algorithms, because of the very strong limitations of local knowledge.
We also designed some interesting algorithms, most notably the Dwork, Lynch, Stockmeyer algorithm for consensus with partial synchrony.

By 1990, distributed computing theory was an active research field.  It included much interesting theory, but also connections with various kinds of distributed systems.  The Principles of Distributed Computing (PODC) conference was well established and active.

Around 1990, I moved from the Distributed Systems group at MIT to the Theoretical Computer Science group; our work had become more theoretical during the previous nine years, diverging from the distributed systems engineering work.  Meanwhile, distributed computing theory had become a recognized sub-discipline within theoretical computer science.  We continued after 1990 with theoretical work on algorithms and impossibility results, and modeling and verification.  Later, we moved to consider different types of distributed systems: wireless networks and mobile systems, and most recently, biological systems.

In the rest of this manuscript, I will describe some of our work since 1990.
This includes additional algorithms and impossibility results in Section~\ref{sec: laterMITalgorithms}, and work on formal models of distributed systems in Section~\ref{sec: laterMITmodels}.  This new work considered not just synchronous and asynchronous systems, but other types of systems, such as timed systems, hybrid continuous/discrete systems, which one might use for distributed software that interacts with an environment, and probabilistic systems. 
I will also mention, in Sections~\ref{sec: wireless} and~\ref{sec: bio}, some recent work on wireless network algorithms and biological distributed algorithms.

\section{Later Work:  Algorithms and Impossibility Results, 1990-2005}
\label{sec: laterMITalgorithms}

During the period 1990-2005, I worked with my students and other collaborators on many algorithms and lower bounds for problems involving communication, consensus, and data consistency.  Timing aspects of algorithms were a major focus.  I have selected a few papers and tried to identify some important themes.  

Also during this period, I wrote my "Distributed Algorithms" textbook~\available{\cite{DBLP:books/mk/Lynch96}}.  It summarized what I thought were the most important ideas of distributed computing theory up to that point.

\subsection{Key publications}

\thepapers{
\noindent
5.1.  Hagit Attiya, Cynthia Dwork, Nancy Lynch, and Larry Stockmeyer. Bounds on the time to reach agreement in the presence of timing uncertainty. Journal of the ACM, 41(1):122--152, January 1994. \\

\noindent
5.2.  Soma Chaudhuri, Maurice Herlihy, Nancy A. Lynch, and Mark R. Tuttle. Tight Bounds for $k$-Set Agreement. Journal of the ACM, pages 47(5):912-943, September, 2000. \\

\noindent
5.3.  Seth Gilbert and Nancy Lynch. Brewer's conjecture and the feasibility of consistent, available, partition-tolerant web services. SIGACT News, 33(2):48-51, June 2002. \\

\noindent
5.4.  Nancy A. Lynch. Distributed Algorithms. Morgan Kaufmann, 1996.
}

\subsection{Communication}

Alan Fekete, Yishay Mansour, and I explored capabilities for communication between two processes on a single channel, as in the \emph{data link} communication layer in the OSI stack of communication layers.  
We proved two results, both giving inherent limitations for this layer of communication.  One showed impossibility of reliable communication if the physical channels could reorder messages---unless we added some extra mechanism such as sequence numbers.  The other showed impossibility of reliable communication in the presence of processor crashes.

Both results appeared in preliminary form in~\available{\cite{DBLP:conf/podc/LynchMF88}}. 
Later journal versions involved other collaborators John Spinelli, Yehuda Afek, Hagit Attiya, Michael Fischer, Da-Wei Wang, and Lenore Zuck, and appeared in~\available{\cite{DBLP:journals/jacm/FeketeLMS93}} and~\available{\cite{DBLP:journals/jacm/AfekAFFLMWZ94}}. 
Again, I/O Automata were used as the underlying model.

\subsection{Timing aspects of algorithms}

An important focus of our algorithmic work during these years involved timing aspects of distributed algorithms.

\paragraph{Time cost of achieving wait-freedom:}
In~\available{\cite{DBLP:journals/jacm/AttiyaLS94}}, we considered the time cost of requiring fault-tolerance in solving the problem of approximate agreement.
This paper recasts the approximate agreement problem from~\available{\cite{DBLP:journals/jacm/DolevLPSW86}} in terms of read/write shared-memory computation. 
The paper contains a collection of results articulating the costs of requiring the strong \emph{wait-free} fault-tolerance property, which says that a process that keeps taking steps is guaranteed to eventually terminate, regardless of the speed or failure of other processes. 

For example, Theorem 7.3 in~\available{\cite{DBLP:journals/jacm/AttiyaLS94}} says that any wait-free algorithm for the $n$-process approximate agreement problem has time complexity at least $\log{n}$. 
The proof considers an execution in which all processes start with $0$, which results in a decision of $0$ by some process $p_i$.  If the schedule is too short, then there is insufficient time for all the processes to "influence" $p_i$'s decision, so there is some other process $p_j$ that does not influence $p_i$.  Then we construct an alternative execution in which $p_j$ starts with $1$ and runs on its own, before anyone else begins.  By the wait-free requirement, $p_j$ must decide on its own, and must decide $1$.  Now running all the other processes as before leads to contradictory decisions.

\paragraph{Mutual exclusion using timing assumptions:}
In~\available{\cite{DBLP:conf/rtss/LynchS92}}, we considered what happens to the number-of-register and time costs of solving the mutual exclusion problem in a shared read/write memory model, when we strengthen the asynchronous model to include some assumptions about timing, specifically, upper and lower bounds on step times.
For this model, we devised an algorithm that guarantees mutual exclusion and deadlock-freedom, using only two shared read/write registers.  This circumvents the lower bound result of~\available{\cite{DBLP:journals/iandc/BurnsL93}}, using timing assumptions.
Its time cost is also low, depending on the parameters describing the timing uncertainty.
The algorithm guarantees its safety property (mutual exclusion), even if the algorithm is run asynchronously, while the liveness property (deadlock-freedom) depends on the timing assumptions.  
We also proved a nearly-matching lower bound tradeoff between the number of registers and the time bound; this proof uses techniques like those in~\available{\cite{DBLP:journals/iandc/BurnsL93}}, extended to the timing-based setting.

The paper~\available{\cite{DBLP:journals/iandc/AttiyaL94}} also considers mutual exclusion in a setting with timing assumptions.
This paper contains algorithms and lower bounds for mutual exclusion in certain centralized and distributed settings, in terms of the parameters describing the timing uncertainty.

A point of possible interest here is that this work was presented as an attempt to begin developing a general theory, with upper and lower bounds, for systems with approximate knowledge of timing.
The underlying shared-memory model used here was expressed in terms of a preliminary type of "Timed I/O Automata" model, basically I/O Automata with added time bounds for the tasks.  We developed more general Timed I/O Automata models later, see Section~\ref{sec: laterMITmodels}.

\paragraph{Consensus:}
Paper 5.1~\available{\cite{DBLP:journals/jacm/AttiyaDLS94}} provides a study of the time costs of reaching consensus in a setting with timing assumptions.
The paper contains new algorithms and lower bounds for consensus in the uncertain-timing setting. 
The key discovery is that the inherent time complexity of consensus in the uncertain-timing setting corresponds essentially to one "long round", whose time cost depends on the timing uncertainty, plus $O(f)$ "short rounds", whose time cost is independent of the timing uncertainty. \\

\thepapers{
\noindent
5.1.  Hagit Attiya, Cynthia Dwork, Nancy Lynch, and Larry Stockmeyer. Bounds on the time to reach agreement in the presence of timing uncertainty. Journal of the ACM, 41(1):122--152, January 1994. \\
}

For the upper bound, we designed an algorithm that uses a simple failure detector, implemented using timeouts. 
Briefly, the algorithm executes alternating rounds, in which processes try to decide $0$ at even-numbered rounds and $1$ and odd-numbered rounds.  Each round is classified as either $quiet$ or $non$-$quiet$; in a $non$-$quiet$ round, every process receives
an explicit message telling it to advance to the next round, whereas in a $quiet$ round, some process fails to receive such a message.  
We show that $quiet$ rounds are short, whereas $non$-$quiet$ rounds may be long, with their time depending on the timing uncertainty.  Occurrence of a $non$-$quiet$ round leads to rapid termination.

For the lower bound, we used an intricate argument that includes a combination of many ingredients from our previous work on impossibility results:  a chain argument as in~\available{\cite{DBLP:journals/ipl/FischerL82}}, plus a bivalence argument as in~\available{\cite{DBLP:journals/jacm/FischerLP85}}, 
plus arguments about shifting, stretching, and shrinking time as in~\available{\cite{DBLP:journals/iandc/LundeliusL84,DBLP:conf/rtss/LynchS92}}. 

\paragraph{Gradient clock synchronization:} 
In~\available{\cite{FanLynch-gradient}}, my PhD student Rui Fan and I introduced the distributed \emph{gradient clock synchronization} problem.
As in traditional distributed clock synchronization, we considered a network of nodes equipped with hardware clocks with bounded drift. Nodes compute logical clock values based on their hardware clocks and message exchanges, and the goal is to synchronize the nodes’ logical clocks as closely as possible. 
The new requirement is that the skew between any two nodes’ logical clocks be bounded by a nondecreasing function of the uncertainty in message delay (distance) between the two nodes. That is, we require nearby nodes to be closely synchronized, and allow faraway nodes to be more loosely synchronized. 
We proved a lower bound of $\Omega(d + \log{D} \log \log D)$ on the worst case clock skew between two nodes at distance $d$ from each other, where
$D$ is the diameter of the network. 
This showed that clock synchronization is not a local property, in the
sense that the clock skew between two nodes depends not only on the distance between the nodes, but also on the size of the entire network.  \\

Our theoretical effort on algorithms and lower bounds for timed systems led us to another, parallel research effort, on developing formal "Timed I/O Automata" models for distributed systems with timing assumptions and guarantees.
We wanted these models to serve as the basis for theoretical work as described here, and also, to be useful for practical system description.
See Section~\ref{sec: laterMITmodels} for more about this modeling work. 

\subsection{$k$-Set Agreement}
We continued work related to fault-tolerant distributed consensus in the basic synchronous model.
Previous results on fault-tolerant consensus in the synchronous model showed that $f+1$ rounds are both necessary and sufficient, over a range of different types of failures.
Paper 5.2~\available{\cite{DBLP:journals/jacm/ChaudhuriHLT00}} generalizes these bounds to the problem of \emph{$k$-set consensus}, which allows $k$ possible decisions to be output, instead of just one.
It considers stopping failures only.
It turns out, perhaps surprisingly, that $\lfloor f/k \rfloor + 1$ rounds are both necessary and sufficient to solve $k$-set consensus for $f$ failures.  Thus, allowing $k$ possible answers reduces the time for ordinary consensus by dividing it by $k$. \\

\thepapers{
\noindent
5.2.  Soma Chaudhuri, Maurice Herlihy, Nancy A. Lynch, and Mark R. Tuttle. Tight bounds for $k$-Set agreement. Journal of the ACM, pages 47(5):912-943, September, 2000.\\
}

The algorithm is straightforward:  every process keeps propagating the minimum value that it has ever seen.  It turns out that the number of distinct minimum values remaining at different processes decreases to at most $k$ by the stated number of rounds.

The lower bound extends the lower bound ideas from~\available{\cite{DBLP:journals/ipl/FischerL82}} to $k$ dimensions, now allowing executions to "morph" gradually between $k+1$ extreme executions in each of which the initial value is fixed at one of the values $0,\ldots,k$, and no failures occur.  
In each execution, we can identify the output values for all the non-failed processes.
If the executions are too short, then we can apply Sperner's lemma from algebraic topology to identify a single execution in the structure that exhibits $k+1$ different output values for non-failed processes. 

\subsection{Data consistency}

We continued working on distributed database issues during this time.  Here are two highlights.

\paragraph{Eventual serializability:}
In~\available{\cite{DBLP:journals/tcs/FeketeGLLS99}}, we presented a new specification for a data service with guarantees that are weaker than the common notion of serializability, together with an algorithm that meets the specification.
Our specification and algorithm are based on earlier work by Ladin et al.~\cite{10.1145/93385.93399}.
We gave a more abstract, general, formal specification of the guarantees of an \emph{eventual serializable} distributed data service.  We gave an algorithm that generalizes the one in~\cite{10.1145/93385.93399} so that it applies to arbitrary data types and accommodates more kinds of ordering constraints.

The basic idea is that the system maintains a partial ordering of operations, which gets refined over time to a total ordering.  Operations may get immediate responses that don't reflect their entire prefix in the final order, but the client may specify constraints on which previous operations must be taken into account.
(This sounds reminiscent of what happens in many blockchain algorithms.)

\paragraph{Consistency, Availability, and Partition-Tolerance (CAP):}
Paper 5.3~\available{\cite{DBLP:journals/sigact/GilbertL02}} is a short paper by my PhD student Seth Gilbert and myself.  It formalized a well-known informal conjecture by Eric Brewer, a leading Berkeley systems researcher, about capabilities of distributed data management systems.
The claim was that it is not possible, in general, to implement web-based data services in such a way as to guarantee all three of the properties of \emph{Consistency}, \emph{Availability}, and \emph{Partition-Tolerance}. 
We expressed this claim formally, in terms of read/write objects distributed in a network of asynchronous processes.  Besides requiring data consistency (atomicity), we required that all operations must terminate, even in the presence of lost messages. 

Our paper was published in the magazine SIGACT News, rather than in a standard journal or conference.
Nevertheless, it has become my third most-referenced publication, exceeded only by my "Distributed Algorithms" book and the FLP paper. 
This may be because the paper is rather simple technically and also because its results say something that is meaningful to distributed systems researchers and developers working on web-based data services. \\

\thepapers{
\noindent
5.3.  Seth Gilbert and Nancy Lynch. Brewer's conjecture and the feasibility of Consistent, Available, Partition-tolerant web services. SIGACT News, 33(2):48-51, June 2002. \\
}

We proved the impossibility result using a familiar type of network-splitting argument. 
%
We then went on to consider the problem using variations on the model:
some involved modifying the timing assumptions, while others involved weakening the consistency guarantees.
Some of these variations maintained impossibility, while others admitted solutions.

This work led to quite a few later papers and much discussion.  I won't try to survey these, but just point to a $10$-years-later issue of IEEE Computer magazine that was largely devoted to perspectives on the CAP theorem~\available{\cite{IEEECAP}}.

\subsection{Distributed Algorithms book}

I also wrote my "Distributed Algorithms" textbook~\available{\cite{DBLP:books/mk/Lynch96}}, Publication 5.4, during this time period.  
It summarized the state of the field at that time, including basic algorithms, key impossibility results, and modeling and proof techniques.
It has proved to be a successful textbook and reference for the field.  
It was based on my course notes from over $10$ years of teaching my MIT graduate Distributed Algorithms course. \\

\thepapers{
\noindent
5.4.  Nancy A. Lynch. Distributed Algorithms. Morgan Kaufmann, 1996.
}

\section{Later Work:  Formal Models and Methods, 1990-2005}
\label{sec: laterMITmodels}

During the years 1990-2005, in addition to continuing our work on algorithms and impossibility results as I described in Section~\ref{sec: laterMITalgorithms}, my collaborators and I engaged in another major research effort, on developing formal models and proof methods for describing and reasoning about distributed systems.  
The need for formal models and methods arose both from our work on theoretical algorithms, and from our attempts to model real distributed systems. 

We already had three major papers on formal models for distributed systems~\available{\cite{DBLP:journals/tcs/LynchF81,DBLP:conf/podc/LynchT87,LynchT89}}.
These contained models for asynchronous distributed systems, with the beginnings of treatment of timing.
The new work involved models for asynchronous and timing-dependent systems, hybrid continuous/discrete systems, and probabilistic systems.

This work got me involved in the formal methods research community, based mainly in Europe, and represented by such conferences as CONCUR, CAV, LICS, and RTSS.
Key collaborators during this time were my postdocs Frits Vaandrager and Dilsun Kirli Kaynar, and my PhD student Roberto Segala.

In this section, I describe our formal methods work during 1990-2005.  in the final Subsection~\ref{sec: verification}, I give some examples of system modeling and verification projects that we carried out during this time, based on our various formal models.

\subsection{Key publications}

\thepapers{
\noindent
6.1.  Nancy Lynch, Roberto Segala, and Frits Vaandrager. Hybrid I/O Automata. Information and Computation, 185(1):105-157, August 2003. \\

\noindent
6.2.  Dilsun Kirli Kaynar, Nancy A. Lynch, Roberto Segala, and Frits W. Vaan-
drager. The Theory of Timed I/O Automata, Second Edition. Synthesis
Lectures on Distributed Computing Theory. Morgan and Claypool Pub-
lishers, 2010.  \\

\noindent
6.3.  Roberto Segala and Nancy Lynch. Probabilistic simulations for probabilistic processes. Nordic Journal of Computing, 2(2):250--273, August 1995.  \\

\noindent
6.4.  Nancy Lynch, Roberto Segala, and Frits Vaandrager. Observing Branching Structure through Probabilistic Contexts. Siam Journal on Computing, 37(4):977-1013, September 2007. \\

\noindent
6.5.  Alan Fekete, Frans Kaashoek and Nancy Lynch. Implementing Sequentially-Consistent Shared Objects Using Group and Point-to-Point Communication . Journal of the ACM, 45(1):35-69, January, 1998. \\

\noindent
6.6.  Alan Fekete, Nancy Lynch, and Alex Shvartsman. Specifying and Using a Partitionable Group Communication Service. ACM Transactions on Computer Systems, 19(2):171-216, May 2001. \\
}

\subsection{Timed (and untimed) system models}

My work with Frits Vaandrager focused on basic automata models for concurrent systems, first for untimed systems, and then for timed systems.

\paragraph{Untimed system models:}
Our initial work on untimed systems resulted in the paper\available{~\cite{DBLP:journals/iandc/LynchV95}}.  
The focus of that work was the use of abstraction mappings of various kinds to reason about the correctness of systems modeled as automata; this general idea was already present in~\available{\cite{DBLP:conf/podc/LynchT87}}. 
The paper~\available{\cite{DBLP:journals/iandc/LynchV95}} contains many new results but can also be read as a comprehensive survey of the area.

The paper covers several different types of mappings, starting with simple \emph{refinements}; a refinement is a (single-valued) function from the state of the lower-level automaton to the state of the higher-level automaton, which preserves step-by-step behavior.

The paper continues with \emph{forward simulations}, which are multivalued mappings that again preserve the step-by-step behavior; now the preservation condition uses existential quantification over the possible choices of the state after the transition.
Forward simulations are more general, more abstract versions of the concrete \emph{history variable} mechanism studied earlier by Owicki and Gries~\cite{DBLP:journals/acta/OwickiG76} and by Abadi and Lamport~\cite{ABADI1991253}.
We continued with \emph{backward simulations}, which again are multivalued mappings, but that use existential quantification over the possible choices of the state before the transition.
Backward simulations are general, abstract versions of the \emph{prophecy variable} mechanism of~\cite{ABADI1991253}.
The paper also contains combinations of forward and backward simulations, which are shown to be complete for expressing implementation relationships between automata.

\paragraph{Timed system models:}
During these years, we worked quite a lot on developing formal models for timed systems, i.e., systems with some timing assumptions.  Our model evolved to its final version~\available{\cite{DBLP:series/synthesis/2010Kaynar}}
in a series of steps.

First, Hagit Attiya and I followed our work in~\available{\cite{DBLP:journals/iandc/AttiyaL94}} with another paper that focused on modeling issues for timed systems~\available{~\cite{DBLP:journals/dc/LynchA92}}. 
There, we defined timed automata in terms of untimed I/O Automata plus "boundmaps" which express upper and lower bounds on the time required for the next steps of particular tasks to occur. 
We intended this as a general, systematic way of incorporating time information into I/O Automata.
By this time, we were convinced of the value of two particular proof methods for untimed automata:  invariant assertions, and abstraction mappings.  
In order to make these methods work for I/O Automata with boundmaps, we developed a systematic way of incorporating the timing information given in boundmaps into the automaton state.
We did this with special new state components representing predictions of when the tasks will next perform a step.  
With this addition to the model, it was easy to adapt invariant and abstraction mapping methods to work in timed systems. 

In~\available{\cite{DBLP:journals/iandc/LynchV96}}, Frits Vaandrager and I generalized the model used in~\available{\cite{DBLP:journals/dc/LynchA92}} further, to allow a timed automaton to have states with arbitrary structure, and to have both discrete transitions and time-passage steps.  
We removed the requirement for the particular "task" structure of I/O Automata.  
This generalization turned out to be able to support the full range of mappings (refinements, forward simulations, backward simulations, etc.) that we presented for untimed systems in~\available{\cite{DBLP:journals/iandc/LynchV95}}.  
A key aspect of these timed automata is their notion of external behaviors:  now they are not just traces, which are sequences of actions, but timed traces, which are sequences of actions with associated times of occurrence.

Frits Vaandrager, Roberto Segala, and I produced a further extension in Paper 6.1~\available{\cite{DBLP:journals/iandc/LynchSV03}}. 
In order to capture the behavior of real-time systems, such as robot-control systems, we extended the model to include continuous behavior.
The combination of discrete and continuous behavior, i.e., \emph{hybrid} behavior, was a growing research direction at the time, represented by the new Hybrid Systems Computation and Control conference.
In addition to time-passage steps as in~\available{\cite{DBLP:journals/iandc/LynchV96}}, we included \emph{trajectories}, which describe the evolution of a state through continuous time.
The resulting model still supports composition and abstraction; the complete theory appears in~\available{\cite{DBLP:journals/iandc/LynchSV03}}. \\

\thepapers{
\noindent
6.1.  Nancy Lynch, Roberto Segala, and Frits Vaandrager. Hybrid I/O Automata. Information and Computation, 185(1):105-157, August 2003. \\
}

Actually, this paper~\available{\cite{DBLP:journals/iandc/LynchSV03}} goes further, allowing not just internal state, but also input and output variables, to evolve continuously.  This provides support for describing continuous interaction between components, in addition to the usual discrete interaction via shared actions.  But this generality led to some technical complications involving composition (now we had to worry about solvability of systems of differential equations).

The final version of our Timed I/O Automata model appears in the monograph~\available{\cite{DBLP:series/synthesis/2010Kaynar}}, Publication 6.2. 
The Timed I/O Automata in this monograph have all the generality of our Hybrid I/O Automata in~\available{\cite{DBLP:journals/iandc/LynchSV03}}, except for the external (input and output) variables.  
Thus, the model includes trajectories describing evolution of the internal state, but the only communication between components is by discrete input and output actions.  
This simplification did not seem to reduce the practical expressive power of the model much, but it did simplify the theory related to composition and, we thought, would make the model easier to use for applications.

The monograph~\available{\cite{DBLP:series/synthesis/2010Kaynar}} contains the complete theory of Timed I/O Automata, including definitions for timed automata and timed system behavior, invariants, and simulation relations.  It contains the theory for the composition and hiding operators, and many simple examples.  \\

\thepapers{
\noindent
6.2.  Dilsun Kirli Kaynar, Nancy A. Lynch, Roberto Segala, and Frits W. Vaan-
drager. The Theory of Timed I/O Automata, Second Edition. Synthesis
Lectures on Distributed Computing Theory. Morgan and Claypool Pub-
lishers, 2010. 
}

\subsection{Probabilistic system models}
\label{sec: prob-models}

Paper 6.3~\available{\cite{DBLP:journals/njc/SegalaL95}} shows how probabilistic systems can also be understood using abstraction mappings. 
The conference version of this appeared in CONCUR in 1994~\available{\cite{DBLP:conf/concur/SegalaL94}}, and recently won a Test-of-Time award from the CONCUR community. \\

\thepapers{
\noindent
6.3.  Roberto Segala and Nancy Lynch. Probabilistic simulations for probabilistic processes. Nordic Journal of Computing, 2(2):250--273, August 1995.  \\
}

To summarize these papers, I give the citation from the CONCUR Test-of-Time award:
\begin{quote}
The paper “Probabilistic simulations for probabilistic processes”, published by Roberto Segala and Nancy Lynch at CONCUR 1994, receives one award for introducing the ‘simple’ probabilistic automata model. Unlike earlier attempts to embrace probabilities, transition targets here are probability distributions over states, and this makes it possible to lift core process algebraic results in a very elegant manner. Probabilistic automata have quickly been recognised as the pivotal link between classical concurrency theory and the theory of discrete-state Markov processes. They have become the central subjects of probabilistic model checking, and are echoed in a range of very influential modelling formalisms including probabilistic timed automata, probabilistic hybrid automata, and Markov automata.
\end{quote}

There is also a published interview of Roberto and myself about this paper, conducted by Luca Aceto~\available{\cite{Aceto-interview}}.

Roberto's 1995 PhD thesis, entitled "Modeling and Verification of Randomized Distributed Real-Time Systems", defined untimed and timed Probabilistic I/O Automata and set out their theory~\available{\cite{Segala95}}.

In the following years, Roberto and I continued working on probabilistic automata, joined at some point by Frits Vaandrager.  A key difficulty arose in our study of composition of probabilistic automata, as a consequence of the fact that our automata can include both probabilistic and nondeterministic choices.
Namely, this combination implies that, for a notion of external behavior to be compositional, it must expose "too much" of the internal branching structure of the automaton.
This is made precise and proved in Paper 6.4~\available{\cite{DBLP:journals/siamcomp/LynchSV07}}.
Roughly speaking, the problem arises when the nondeterministic choices get resolved by an adversarial "scheduler" that makes unfavorable choices based on knowledge of internal aspects of the prior execution. \\

\thepapers{
\noindent
6.4.  Nancy Lynch, Roberto Segala, and Frits Vaandrager. Observing Branching Structure through Probabilistic Contexts. Siam Journal on Computing, 37(4):977-1013, September 2007. \\
}

To get around the problem demonstrated in~\available{\cite{DBLP:journals/siamcomp/LynchSV07}}, it seems necessary to restrict the power of the adversarial scheduler that controls the nondeterministic choices within a probabilistic automaton.
Such restrictions have appeared in some of our subsequent work, such as~\available{\cite{DBLP:journals/tcs/CheungLSV06}} and~\available{\cite{CCKLLPS-jcss}}.
In~\available{\cite{DBLP:journals/tcs/CheungLSV06}}, we proposed a restricted scheduler based on a predetermined schedule of components and an adversarial scheduler within components.
In~\available{\cite{CCKLLPS-jcss}}, we described a framework for modeling and verifying security protocols, with a restricted scheduler that is based on an oblivious, global scheduler for tasks (a concept borrowed from I/O Automata), coupled with a local scheduler for resolving nondeterminism among the choices of actions within a task.
We called the resulting model \emph{Task-Structured Probabilistic I/O Automata}.

Other types of restricted schedulers also arise in some of our more recent work on wireless network communication and neural networks. 

\subsection{Verification examples}
\label{sec: verification}

We used our formal methods work as the basis for many projects on modeling and verifying distributed systems, including distributed data processing systems, real-time control systems, and security protocols. Here is a brief overview, broken down according to the type of model used.

\paragraph{Asynchronous systems:} 
The nested transaction work that I discussed in Section~\ref{sec: earlyMIT} was an extended case study, presenting many algorithms based on ones that were developed in the distributed systems community (generalized in some cases).  We put everything in a uniform framework, in terms of I/O Automata.  We published a comprehensive book summarizing the entire modeling project~\available{\cite{DBLP:books/mk/LynchMWF93}}.  
We also modeled and verified other distributed systems algorithms, and in some cases, found errors in the algorithms and helped to fix them.  I describe a few examples here.

First, in Paper 6.5~\available{\cite{DBLP:journals/jacm/FeketeKL98}}, my PhD student Alan Fekete and I worked with Frans Kaashoek to try to understand, in formal terms, an interesting distributed algorithm that Frans had helped to develop for the Orca shared-object system, at the Vrije University in Amsterdam.  \\

\thepapers{
\noindent
6.5.  Alan Fekete, Frans Kaashoek and Nancy Lynch. Implementing Sequentially-Consistent Shared Objects Using Group and Point-to-Point Communication . Journal of the ACM, 45(1):35-69, January, 1998. \\
}

The algorithm implements a sequentially consistent collection of shared read/update objects in an asynchronous distributed network.
It caches objects in the local memory of processors according to application needs.  Each read operation accesses a single copy of the object, while each update accesses all copies. A process uses broadcast communication when it sends messages to all the copies of an object, and uses point-to-point communication when it sends a message to a single copy, and when it returns a reply. 
Copies of an object are kept consistent using a strategy based on sequence numbers for broadcasts.

We presented the algorithm using two layers of abstraction. The lower layer uses the given broadcast and point-to point communication services, plus sequence numbers, to implement a new, powerful communication service called a \emph{context multicast channel}. The higher layer uses the context multicast service to manage the object replication consistently. 
We described and verified both layers and their combination using I/O Automata and abstraction methods.

Actually, while attempting to verify our model of the existing Orca algorithm, we found a logical error in the implemented algorithm, namely, it omitted sequence numbers from some messages that needed to include them.  We produced a corrected version of the algorithm and verified that.  The corrected algorithm was then incorporated into the Orca system. 

Second, in Paper 6.6~\available{\cite{DBLP:journals/tocs/FeketeLS01}}, Alan Fekete, Alex Shvartsman, and I provided a new formal specification of a \emph{partitionable Group Communication Service (GCS)}.  GCSs were a topic of great interest in the distributed systems research community in the late 1990s, as building blocks for fault-tolerant distributed systems.  A GCS  is based on a \emph{group membership service}, which maintains changing versions of the set of members of the group.
The GCS then manages communication among the current group members.  \\

\thepapers{
\noindent
6.6.  Alan Fekete, Nancy Lynch, and Alex Shvartsman. Specifying and Using a Partitionable Group Communication Service. ACM Transactions on Computer Systems, 19(2):171-216, May 2001. \\
}

Initially, it was not completely clear to us what the formal specification for the GCS should be.
To pin down the details, we developed our specification for the GCS in the context of a particular application:  an \emph{ordered broadcast} service. Specifically, we designed an algorithm to implement ordered broadcast over a GCS, and proved the algorithm to be correct, based on our new formal specification of the GCS. This work served to clarify some notions that had been discussed less formally in the systems research community.

I also collaborated with Ken Birman and Jason Hickey at Cornell, with the aim of understanding the behavior of their Ensemble GCS, in formal terms.
Ensemble was then a widely used GCS that supported distributed programming by providing precise guarantees for synchronization, message ordering, and message delivery.  
We used I/O Automata again, to formalize, specify, and try to verify the correctness of the Ensemble implementation~\available{\cite{DBLP:conf/tacas/HickeyLR99}}.
Once again, the verification effort uncovered an algorithmic error in the implementation, which was subsequently repaired in the implementation.

\paragraph{Timed and hybrid systems:}
We used our timed automata and hybrid automata models extensively, for verifying algorithms derived from practical real-time systems.
Here I mention a few examples.

First, in the early 1990s, I worked with Butler Lampson and my PhD student Jorgen Sogaard-Anderson on modeling and verifying two \emph{At-Most-Once message delivery} protocols.  These were patterned, respectively, after the standard protocol for setting up network connections used in TCP and other transport protocols, and a clock-based protocol of Liskov, Shrira, and Wroclawski~\cite{LiskovBarbara1991Eamb}.  We specified the algorithms and verified their key properties, using our Timed I/O Automata model.  The proofs used invariants and abstraction mappings. 

Second, I worked with Prof. Shankar Sastry and members of his Berkeley hybrid systems group, in particular, his PhD student John Lygeros, on modeling the behavior of automated vehicle systems.   In 1998, we carried out a project modeling strings of automated vehicles and proving certain safety properties for their behavior~\available{\cite{10.1007/3-540-64358-3_45}}.  
We identified an emergency vehicle maneuver---one in which one car brakes to decelerate as fast as possible---and identified conditions under which crashes are guaranteed to be avoided.  This work used our Hybrid I/O Automata model, and inductive proof methods, taking into account both discrete and continuous transitions.  We also identified conditions under which crashes may occur; these proofs are by construction.

Third, in a project involving John Lygeros and my PhD student Carolos Livadas, we developed a model for the behavior of the recently-introduced aircraft \emph{Traffic alert and Collision Avoidance System (TCAS)}. 
In our paper\available{~\cite{871302}}, we provided an abstract model of the essential behavior of the main TCAS collision avoidance algorithm, and proved that it in fact works to avoid aircraft collisions.
Our work again used our Hybrid I/O Automata model.
We presented this work to TCAS developers at Lincoln Labs.
Our approach complemented the detailed code discussions that were the norm at that time for developers of such safety-critical software.
We advocated for the use of abstract modeling and formal verification, in addition to detailed code discussions.

Fourth, in 2002, my PhD student Sayan Mitra and I carried out a project with Prof. Eric Feron in MIT's Aero-Astro department and his student~\available{\cite{10.1007/3-540-36580-X_26}}.  This involved modeling a toy helicopter system used for instructional purposes and verifying its safety properties.  Again, we used Hybrid I/O Automata.

Finally, in a little-known, simple paper~\available{\cite{Lynch-3level}}, I showed two ways in which abstraction mappings could be used for real-time systems modeled using Hybrid I/O Automata:  for expressing the relationship between a derivative-based description of a system and an explicit functional description, and to express the relationship between a system description in which corrections are made at discrete sampling points and a version in which corrections are made continuously.  Both types of relationships can be proved with simple forward simulation relations. 

\paragraph{Probabilistic systems:}
Our main case study here involved verification of basic security protocols using \emph{Task-Structured Probabilistic I/O Automata}~\available{\cite{CCKLLPS-jcss}}.
Specifically, using Task-Structured Probabilistic I/O Automata, we modeled and verified some basic security protocols following Ran Canetti's \emph{Universal Composability} approach.
For details, see, for example,~\available{~\cite{DBLP:journals/deds/CanettiCKLLPS08}}.
\\
\\
In Sections~\ref{sec: laterMITalgorithms} and~\ref{sec: laterMITmodels}, I have summarized our contributions during 1990-2005 to the continuing active development of the field of distributed computing theory.  Our work included theoretical algorithms and lower bounds, new formal models and methods for reasoning about distributed algorithms and systems, and applications.
This work was part of the very active development of the field of distributed computing theory during these years.

Our main contributions during this time included some new of the new theoretical algorithms and lower bounds, including various results involving timing in distributed algorithms.  They also included applications of the theory to understand issues arising in the distributed systems community, such as Brewer's CAP conjecture, 
They included new formal models for timed, hybrid, and probabilistic distributed systems, and applications of these models to verify systems algorithms.
They also included a major textbook for the field, 

So far, I have described work on algorithms related to traditional distributed systems, including timed and hybrid systems.  After 2005, our focus shifted to new types of distributed systems:  wireless networks and biological systems.  In these systems, noise, uncertainty, and change predominate, and so the systems must be flexible, robust, and adaptive.  We continued carrying out the same general kinds of research, namely, defining problems to be solved in these systems, developing algorithms, proving correctness and performance properties, and defining formal foundations.

\section{Still Later Work:  Wireless Network Algorithms, 2005 and Later}
\label{sec: wireless}

We continued working on algorithms and lower bounds after 2005, but now considering wireless systems, including mobile wireless systems.  
We focused on ad hoc wireless networks, with no central base station.
These might be used, for example, by soldiers in hostile territory, first responders in a disaster area, or swarms of robots.
The main challenge was to design new algorithms for important problems (communication, searching, maintaining data, coordinating agents).
We also wanted to define good abstraction layers to mask some of the difficulties and make it easier to develop applications for ad hoc wireless networks.

\subsection{Key publications}

\thepapers{
\noindent
7.1.  Seth Gilbert, Nancy Lynch, and Alex Shvartsman. RAMBO: A Robust, Reconfigurable Atomic Memory Service for Dynamic Networks. Distributed Computing, 23(4):225-272, December 2010. This is a slightly corrected version of the journal version. \\

\noindent
7.2.  Matthew Brown, Seth Gilbert, Nancy Lynch, Calvin Newport, Tina Nolte, and Michael Spindel. The Virtual Node Layer: A Programming Abstraction for Wireless Sensor Networks. ACM SIGBED Review, 4(3), July 2007.  Also, Proceedings of the the International Workshop on Wireless Sensor Network Architecture (WWSNA), Cambridge, MA, April, 2007. \\

\noindent
7.3.  Seth Gilbert, Nancy Lynch, Sayan Mitra, and Tina Nolte. Self-Stabilizing Robot Formations Over Unreliable Networks. ACM Transactions on Autonomous and Adaptive Systems, 4(3):17.2-17.27, July 2009.\\

\noindent
7.4.  Fabian Kuhn, Nancy Lynch and Rotem Oshman. Distributed Computation in Dynamic Networks. Proceedings of the 42nd ACM Symposium on Theory of Computing (STOC 2010), Cambridge, MA, pages 513-522, June 2010. \\

\noindent
7.5.  Alejandro Cornejo, Fabian Kuhn, Ruy Ley-Wild, and Nancy Lynch. Keeping Mobile Robot Swarms Connected. In Idit Keidar, editor, Distributed Computing, DISC 2009: 23rd International Symposium on Distributed Computing, Elche/Elx, Spain, September 23-25 2009, volume 5805 of Lecture Notes in Computer Science, pages 496-511, 2009. Springer. \\
}

\subsection{Reconfigurable atomic memory}

In 2002, we developed the \emph{RAMBO (Reconfigurable Atomic Memory for Basic Objects)} algorithm for maintaining the appearance of globally-shared read/write memory in changing networks of mobile devices~\available{\cite{RAMBO}}. 
This might be used, for example, by a company of soldiers operating in a hostile
environment, without access to fixed wireless infrastructure.\footnote{For me, this work was inspired by start of the war in Afghanistan, soon after the events of September 11, 2001.  I imagined the difficulties that soldiers roaming in an unfamiliar area would have in communicating, and especially, in maintaining reliable data.  Of course, no cell towers would be available, so everything needed to be implemented on the mobile devices themselves.} 

The algorithm replicated each object at a set of nodes.
Read and write operations were implemented by accessing read quorums and write quorums of copies, respectively.
That was sufficient for relatively stable situations, in which only small, transient changes occurred.
For larger and more permanent changes, the algorithm also supported explicit \emph{reconfiguration} operations.
To move from an old configuration to a new one, the algorithm used the simple trick of employing both configurations for a while, during the period when the configuration was changing.

The final version of the paper appeared in 2010, in Paper 7.1~\available{\cite{DBLP:journals/dc/GilbertLS10}}, after several earlier versions. \\

\thepapers{
\noindent
7.1.  Seth Gilbert, Nancy Lynch, and Alex Shvartsman. RAMBO: A Robust, Reconfigurable Atomic Memory Service for Dynamic Networks. Distributed Computing, 23(4):225-272, December 2010. This is a slightly corrected version of the journal version. \\
}

From the abstract:

\begin{quote}
In this paper, we present Rambo, an algorithm for emulating a read/write distributed shared memory in a dynamic, rapidly changing environment. Rambo provides a highly reliable, highly available service, even as participants join, leave, and fail. In fact, the entire set of participants may change during an execution, as the initial devices depart and are replaced by a new set of devices. Even so, Rambo ensures that data stored in the distributed shared memory remains available and consistent. 

There are two basic techniques used by Rambo to tolerate dynamic changes. Over short intervals of time, replication suffices to provide fault-tolerance. While some devices may fail and leave, the data remains available at other replicas. Over longer intervals of time, Rambo copes with changing participants via reconfiguration, which incorporates newly joined devices while excluding devices that have departed or failed. The main novelty of Rambo lies in the combination of an efficient reconfiguration mechanism with a quorum-based replication strategy for read/write shared memory. 

The Rambo algorithm can tolerate a wide variety of aberrant behavior, including lost and delayed messages, participants with unsynchronized clocks, and, more generally, arbitrary asynchrony. Despite such behavior, Rambo guarantees that its data is stored consistently. We analyze the performance of Rambo during periods when the system is relatively well-behaved: messages are delivered in a timely fashion, reconfiguration is not too frequent, etc. We show that in these circumstances, read and write operations are efficient, completing in at most eight message delays.
\end{quote}


\subsection{Virtual Nodes}

We introduced the concept of \emph{Virtual Nodes (VNs)}, as an abstraction for building applications over mobile wireless networks.  
The idea is to allow the rather chaotic collection of mobile nodes to cooperate to implement more reliable and stable abstract state machines called Virtual Nodes, which could then be programmed as if they were real machines.\footnote{Virtual Nodes are reminiscent of the Virtual Supervisor abstraction that we used in our very first paper on distributed computing theory~\available{\cite{BurnsJLFP82}}.  That was for a shared memory model.}
Our initial efforts focused on Virtual Mobile Nodes\available{~\cite{DBLP:conf/wdag/DolevGLSSW04}}, but we later emphasized Virtual Stationary Nodes at fixed geographical locations.

This work led to many papers including~\available{\cite{DBLP:conf/wdag/DolevGLSSW04,DBLP:journals/dc/DolevGLSW05,DGLLN-opodis05,DLLN,DBLP:journals/sigbed/BrownGLNNS07, NolteLynch-sss07, NolteLynch-icdcs07, GLMN-sss08}}, PhD theses for Tina Nolte~\available{\cite{Nolte}} and Seth Gilbert\available{~\cite{Gilbert}}, and MEng theses for Matt Brown~\available{\cite{Brown}} and Mike Spindel~\available{\cite{Spindel-meng}}.
I highlight three papers here: an early paper~\available{\cite{DBLP:journals/dc/DolevGLSW05}} on using VNs to implement atomic memory in mobile networks, a basic position paper~\available{\cite{DBLP:journals/sigbed/BrownGLNNS07}} summarizing the general Virtual Node approach, and an application of VNs to robot swarm motion coordination~\available{\cite{DBLP:journals/taas/GilbertLMN09}}. 

First, the paper~\available{\cite{DBLP:journals/dc/DolevGLSW05}} describes a way in which (reliable) mobile nodes can implement read/write atomic memory.  The idea is simply to have the mobile nodes implement Virtual Stationary Nodes at known, fixed  locations, and to have the Virtual Stationary Nodes implement a quorum-based atomic read/write memory algorithm.  
In this setting, a VN might fail, when its local area becomes empty of real mobile nodes.  The paper discusses limited reconfiguration mechanisms, which can be used to bring a failed VN up-to-date when its local area becomes repopulated.

Paper 7.2~\available{\cite{DBLP:journals/sigbed/BrownGLNNS07}} is a short position paper motivating and describing the Virtual Node approach.  It contains an interesting example of \emph{Virtual Traffic Lights}, which could be implemented by computers on the cars. Actually, that was not practical at the time this paper was written since not all cars then had on-board computers, but now this would be quite feasible.\\

\thepapers{
\noindent
7.2.  Matthew Brown, Seth Gilbert, Nancy Lynch, Calvin Newport, Tina Nolte, and Michael Spindel. The Virtual Node Layer: A Programming Abstraction for Wireless Sensor Networks. ACM SIGBED Review, 4(3), July 2007.  Also, Proceedings of the the International Workshop on Wireless Sensor Network Architecture (WWSNA), Cambridge, MA, April, 2007. \\
}

Finally, Paper 7.3~\available{\cite{DBLP:journals/taas/GilbertLMN09}} gives an example of an application of Virtual Stationary Nodes to robot swarm motion coordination, for example, guiding robots to surround a hazardous waste spilll. \\

\thepapers{
\noindent
7.3.  Seth Gilbert, Nancy Lynch, Sayan Mitra, and Tina Nolte. Self-Stabilizing Robot Formations Over Unreliable Networks. ACM Transactions on Autonomous and Adaptive Systems, 4(3):17.2-17.27, July 2009. \\
 }

Other notable papers describe uses of Virtual Nodes to assist in message routing and to implement Virtual Air-Traffic Controllers.

\subsection{Computing in dynamic networks}

In this subsection and the next, I describe two more projects involving computing in mobile networks.
Paper 7.4~\available{\cite{DBLP:conf/stoc/KuhnLO10}} is a theoretical paper by postdoc Fabian Kuhn, PhD student Rotem Oshman, and myself, containing algorithms and lower bounds for dynamic networks in which the network topology changes from round to round. \\

\thepapers{
\noindent
7.4.  Fabian Kuhn, Nancy Lynch and Rotem Oshman. Distributed Computation in Dynamic Networks. Proceedings of the 42nd ACM Symposium on Theory of Computing (STOC 2010), Cambridge, MA, pages 513-522, June 2010.
} \\

The paper assumes a worst-case model in which the communication links for each round are chosen by an adversary, and nodes do not know who their neighbors for the current round are before they broadcast their messages.   We require correctness and termination even in networks that change continually.
Our results rely on a network connectivity property that we called
\emph{$T$-interval connectivity}, which says that, in every $T$ consecutive rounds, there is a stable connected spanning subgraph. 

For this model, the paper contains algorithms by which the nodes can determine the size of the network, and compute any function of their initial inputs, in $O(n^2/T)$ rounds.
It also contains lower bounds for the token dissemination problem, which requires the nodes to disseminate information to all the nodes in the network.

\subsection{Robot swarm algorithms}

PhD student Alejandro Cornejo led a project on algorithms for various motion-planning tasks in robot swarms; see, for example,~\available{\cite{CKLL-disc09,CornejoLynch-icra10,CornejoLynch-opodis10}}.
In particular, Paper 7.5~\available{\cite{CKLL-disc09}} contains the design for a \emph{connectivity service} that can adapt any robot swarm motion planner to ensure that the swarm remains globally connected for communication throughout the computation.
The connectivity service does not interfere with progress of the robots. \\

\thepapers{
\noindent
7.5.  Alejandro Cornejo, Fabian Kuhn, Ruy Ley-Wild, and Nancy Lynch. Keeping Mobile Robot Swarms Connected. In Idit Keidar, editor, Distributed Computing, DISC 2009: 23rd International Symposium on Distributed Computing, Elche/Elx, Spain, September 23-25 2009, volume 5805 of Lecture Notes in Computer Science, pages 496-511, 2009. Springer. 
} \\

We also have recent work on robot swarm computations, but it is a bit too preliminary to discuss here.

\subsection{The Dual Graph model}

All of the work I have described so far for wireless networks assumes that communication is reliable. 
We also considered unreliable communication, formalized in terms of a static \emph{Dual Graph} model.  A Dual Graph consists of a graph $G$ of edges that support reliable communication and a super-graph $G'$ containing additional edges over which messages might or might not be delivered.
The options for delivery are assumed to be controlled by an adversary.
In addition,  the networks include another type of communication unreliability:  message collisions in which colliding messages are lost.

For this difficult setting, we produced many papers containing upper and lower bounds for solving problems such as local and global broadcast, for example,~\available{\cite{DBLP:conf/podc/KuhnLNOR10, DBLP:conf/wdag/GhaffariHLN12, DBLP:conf/podc/GhaffariLN13, CGKLN-jour, DBLP:conf/podc/LynchN15, DBLP:journals/tcs/GilbertLNP20}}. 

Generally speaking, the Dual Graph model was so difficult that we ended up proving mostly lower bounds on the time costs for solving problems.  These lower bounds served to establish a difference in power between wireless models with reliable and unreliable communication.  
We also tried to design algorithms for the Dual Graph model that worked as well as possible, but their performance was not great.

One reason why the algorithms for the Dual Graph model did not perform too well is that they included a combination of probabilistic and nondeterministic choices:  the algorithms are probabilistic, and message delivery is nondeterministic, potentially controlled by an adversary.
As I noted in Section~\ref{sec: prob-models}, that can cause technical problems with regard to composition.  In the case of typical wireless network algorithms, like backoff algorithms, it also causes efficiency problems; see~\available{\cite{DBLP:journals/tcs/GilbertLNP20}}.  
Our way around the problem in~\available{\cite{DBLP:journals/tcs/GilbertLNP20}} is, again, to weaken the power of the adversary.  In this case, we assume that the adversary is stochastic rather than worst-case.  
\\
\\
Thus, our contributions to the wireless network area mainly involved new algorithms and abstractions to try to make programming of such networks tractable.
In the remaining section of this paper, I will briefly discuss our recent work on biological distributed algorithms.

\section{Biological Distributed Algorithms, 2012 and later}
\label{sec: bio}

During the last 12 years or so, my group has been focusing on very different kinds of distributed algorithms, namely, those arising in biological systems.
Most biological systems are distributed---think of colonies of insects, and networks of cells such as brain networks---so it is natural to view them in terms of distributed algorithms.
But they are distributed algorithms of a particular kind:  flexible, robust, and adaptive. 

Biological systems, of course, include many biological details.
However, we expect that less detailed, abstract models of these systems can be defined, and can be treated as abstract distributed algorithms.
We hope that some of the algorithm design and analysis methods from traditional distributed computing theory will carry over to these new kinds of distributed systems.
This could contribute a new type of understanding to the field of biology.
We also hope that, once we understand how the biological mechanisms work, we will be able to adapt them for use in engineered systems such as robot swarms and neural networks.

My group and I have mainly worked on two kinds of biological systems.  The work is still in progress, and preliminary, so I will not say too much about it now.

\subsection{Insect colonies}

Our project on insect colony algorithms has involved designing algorithms that capture key aspects of real insect colony behavior, modeling the algorithms formally, simulating them to try to infer properties of their behavior, and in some cases, following the simulation results with analysis. 

The main problems that we have considered so far include searching for food, reaching consensus on a new nest ("house-hunting")~\available{\cite{DBLP:conf/podc/GhaffariMRL15, DBLP:journals/jcb/ZhaoLP22}}, task allocation~\available{\cite{DBLP:conf/wdag/CornejoDLN14,DBLP:journals/ploscb/RadevaDLNS17,DBLP:conf/sss/SuSDL17, DBLP:conf/spaa/DornhausLMPR20}}, and density estimation~\available{\cite{DBLP:journals/pnas/MuscoSL17, DBLP:journals/corr/MuscoSL16}}.

Generally, we have been successful at modeling these types of insect colony behavior abstractly, and have observed and proved properties that are consistent with experimental observations by insect biologists.  We have made a few observations from our modeling work that might suggest new experiments.

\subsection{Brain networks}

The main theme of our work on brain networks is to model mechanisms that might be used in the brain, for solving various brain problems.  
We have treated these mechanisms as abstract distributed algorithms and have proved properties of their behavior.

Some of our representative papers consider the \emph{Winner-Take-All} problem~\available{\cite{DBLP:journals/corr/abs-1904-12591}}, concept representation and memorization~\available{\cite{DBLP:conf/innovations/HitronLMP20}}, and representation and learning of hierarchically-structured concepts~\available{\cite{DBLP:journals/nn/LynchM21, DBLP:conf/sirocco/LynchM23}}. %
We also developed a basic \emph{formal modeling framework} for brain network algorithms~\available{\cite{DBLP:conf/birthday/LynchM22}}, based on synchronous, stochastic \emph{Spiking Neural Networks (SNNs)}.

\section{Conclusions}

I have described how my collaborators and I helped to develop a theory for distributed systems, in order to understand their capabilities and limitations.
Our contributions have included:
\begin{itemize}
    \item  New distributed algorithms.
    \item  Formal models for distributed algorithms and systems, rigorous proofs and analysis.
    \item  Discovery of errors in existing algorithms and systems.
    \item  Lower bounds and other impossibility results, expressing inherent limitations.
    \item  General modeling and proof methods. 
    \item  Applications to distributed data-management systems, wired and wireless communication systems, real-time systems, and biological systems.
\end{itemize}
Distributed computing theory is now a healthy, active branch of theoretical computer science, and I am happy that we have played some role in its development.


\bibliography{Lifetime}
\end{document}